\documentclass[notitlepage, reprint, superscriptaddress, aps, prb, floatfix, superscriptaddress]{revtex4-2}
\pdfoutput=1

\usepackage[colorlinks=true,citecolor=blue, linkcolor=blue,urlcolor=blue]{hyperref}
\usepackage{titlesec}
\usepackage[final]{pdfpages}
\titleformat{\section} 
{\large\bfseries\color{black}}
{\thesection}{1em}{}
\usepackage{amsmath, mathtools, amssymb, bigints}
\usepackage{hyperref}
\usepackage[normalem]{ulem}
\usepackage{graphicx}
\usepackage{bm}
\usepackage{color}
\usepackage{bbold}

\newcommand{\BM}{\text{BM}}
\newcommand{\SM}{\text{SM}}
\newcommand{\TIM}{\text{TIM}}

\makeatletter
\AtBeginDocument{\let\LS@rot\@undefined}
\makeatother

\begin{document}

\title{Universal Theory of Incoherent Metals}

\author{Aaron Kleger}
\email{Aaron.R.Kleger.gr@dartmouth.edu}

\author{Nikolay V. Gnezdilov}
\email{Nikolay.Gnezdilov@dartmouth.edu}

\author{Rufus Boyack}
\email{Rufus.Boyack@dartmouth.edu}

\affiliation{Department of Physics and Astronomy, Dartmouth College, Hanover, New Hampshire 03755, USA}

\begin{abstract}
Numerous unconventional superconductors such as cuprates, heavy-fermions, and twisted-bilayer graphene exhibit incoherent metallic transport above the superconducting critical temperature. This phenomenon cannot be described with Fermi-liquid theory and has presented a significant theoretical challenge to overcome. We utilize the two-dimensional Yukawa-SYK model of fermions with spatially random coupling to quantum-critical bosons to study transport in a manner which is non-perturbative in the coupling strength. Our work provides a microscopic model of quantum-critical incoherent metals and their concomitant properties, including a non-Boltzmann transport formula between resistivity and quasi-particle lifetime, violation of the Mott-Ioffe-Regel resistivity bound, and violation of the Kovtun-Son-Starinets shear viscosity to entropy density bound.
\end{abstract}

\maketitle

{\it Introduction.---} Over the past three decades, many strongly correlated materials such as cuprate~\cite{Gurvitch1987, Takagi1992, Shen1995, Cooper2009, Ayres2021}, heavy-fermion~\cite{Lohneysen1996, Trovarelli2000, Bruin2013}, and iron-pnictide~\cite{DoironLeyraud2009} superconductors, and more recently twisted bilayer graphene~\cite{Polshyn2019, Wu2019} have been found to exhibit universal features in their thermodynamics and transport properties. The normal phases of these `strange metals'~\cite{Orenstein2000, McGreevy2010, Cha2020, Phillips2022, Patel2023, Savitsky2025} exhibit distinctly non-Fermi-liquid physics~\cite{Lohneysen2007}, such as linear-in-$T$ resistivity~\cite{Varma2020} at low temperatures. An unconventional superconductor has many rich and distinct normal phases~\cite{Keimer2015}, and one of the foremost pressing challenges is to understand the high-temperature regime, which extends beyond the Mott-Ioffe-Regel limit~\cite{Ioffe1960, Mott1972, Hussey2004}. Such a regime is broadly known as an incoherent or bad metal. The challenges in understanding these diverse materials include the absence of a small perturbation parameter and the lack of a quasi-particle description. In this paper, we provide a theoretical model that encapsulates universal features of incoherent metals including: (1) non-Boltzmann transport, (2) resistivity above the Mott-Ioffe-Regel bound, and (3) the possibility of violating the conjectured shear viscosity to entropy density bound~\cite{Kovtun2003, Kovtun2005}. The importance of this work is two-fold: (1) we provide universal predictions, and (2) the framework is not restricted to weak-coupling. Indeed, we find bad-metal physics in the strong-coupling regime. Our work provides the first step to a holistic understanding of the normal phases of unconventional superconductors.

The notions of an incoherent metal~\cite{Jarrell1996, Parcollet1999, Wolfle2002, Hartnoll2015, Davison2015b} and a bad metal (BM)~\cite{Emery1995, Gunnarsson2003, Deng2013, Kokalj2017, Delacretaz2017, Lucas2017, Mousatov2019, Pustogow2021, Valentinis2026} have been proposed to describe systems where the interactions are so strong that fermionic excitations become incoherent, and consequently are not long-lived quasiparticles. Such behavior has been argued~\cite{Wolfle2002} to be quite natural in the regime of high temperatures, where thermal fluctuations are non-negligible. Experimental observation of incoherent behavior has been reported for nickelates~\cite{Jaramillo2014}, cuprates~\cite{Shen2019}, and ultra cold Fermi gases~\cite{Xu2019, Brown2019}. Theoretical studies of incoherent-metals~\cite{Patel2018b, Chowdhury2020} and bad-metals~\cite{Valentinis2026} have been carried out in the context of variants of the Sachdev-Ye-Kitaev (SYK) model~\cite{Ye1993}, which we also implement.

For incoherent metals, the electronic bandwidth $\Lambda$ and Fermi-energy $\varepsilon_F$ are the relevant energy scales in which to meaningfully determine whether the excitation lifetime $\tau$ is short or long. For conventional metals with long-lived quasi-particles, the excitation lifetime satisfies $\Lambda\tau\gg\hbar$, and thus the electronic bandwidth is effectively infinite and transport is well described by semi-classical Drude theory. The case where the transport scattering times are so short that the mean free path is comparable to the interatomic lattice spacing defines the Mott-Ioffe-Regel (MIR) limit~\footnote{Alternative definitions such as $l\sim1/k_F$ or $l\sim2\pi/k_F$ (where $l$ denotes the electronic mean free path and $k_F$ the Fermi momentum) have been employed~\cite{Hussey2004}.}, where Drude-Boltzmann transport theory breaks down. In this regime, $\Lambda\tau\lesssim\hbar$ and metallic transport is highly incoherent; such materials are often referred to as `bad metals' and have a resistivity greater than the quantum unit $\rho>h/e^2$ (given in two spatial dimensions)~\cite{Emery1995, Gunnarsson2003, Mousatov2019}.

The absence of an analytical description of transport in the incoherent-metal regime presents a significant obstacle in the interpretation of experimental data. The most notable example, a `Planckian' scattering rate $1/\tau\sim k_B T/\hbar$~\cite{Hartnoll2022}, is commonly inferred from the $T$-linear resistivity observed in unconventional superconductors~\cite{Legros2019}. However, this inference assumes that resistivity follows a Drude-like relation $\rho\sim1/\tau$. At high temperatures, the resistivity is beyond the MIR limit, and thus a Drude-like transport relation is clearly inapplicable, which calls into question the notion of Planckian dissipation in bad metals~\cite{Fratini2026}. Developing a non-Boltzmann transport relation in the bad-metal regime of an electron-boson model is one of the significant achievements of this paper.

Important characterizations of the absence of a quasi-particle picture have been made on the basis of universal bounds of transport observables. In particular, Kovtun, Son, and Starinets (KSS) have conjectured a lower bound on the shear viscosity to entropy density ratio $\eta/s\geq \hbar/(4\pi k_B)$~\cite{Kovtun2003,Kovtun2005}, while Hartnoll has conjectured a lower bound on diffusivity $D\geq \hbar v_F^2/(k_BT)~\cite{Hartnoll2015}$. For resistivity well below the MIR bound, where $\Lambda\tau\gg\hbar$, the Yukawa-Sachdev-Ye-Kitaev (Y-SYK) model has shown great success in describing a variety of properties of strange metals, such as $T$-linear resistivity and $T\log(1/T)$ heat capacity, as well as hosting superconductivity~\cite{EsterlisSchmalian2019, Wang2020, WangChubukov2020, Patel2023, Valentinis2023, Valentinis2023b, Li2024, Esterlis2025, Kryhin2025, Pan2021, Gnezdilov2025}. 

\begin{figure}[t]
\centering\includegraphics[width=1\columnwidth]{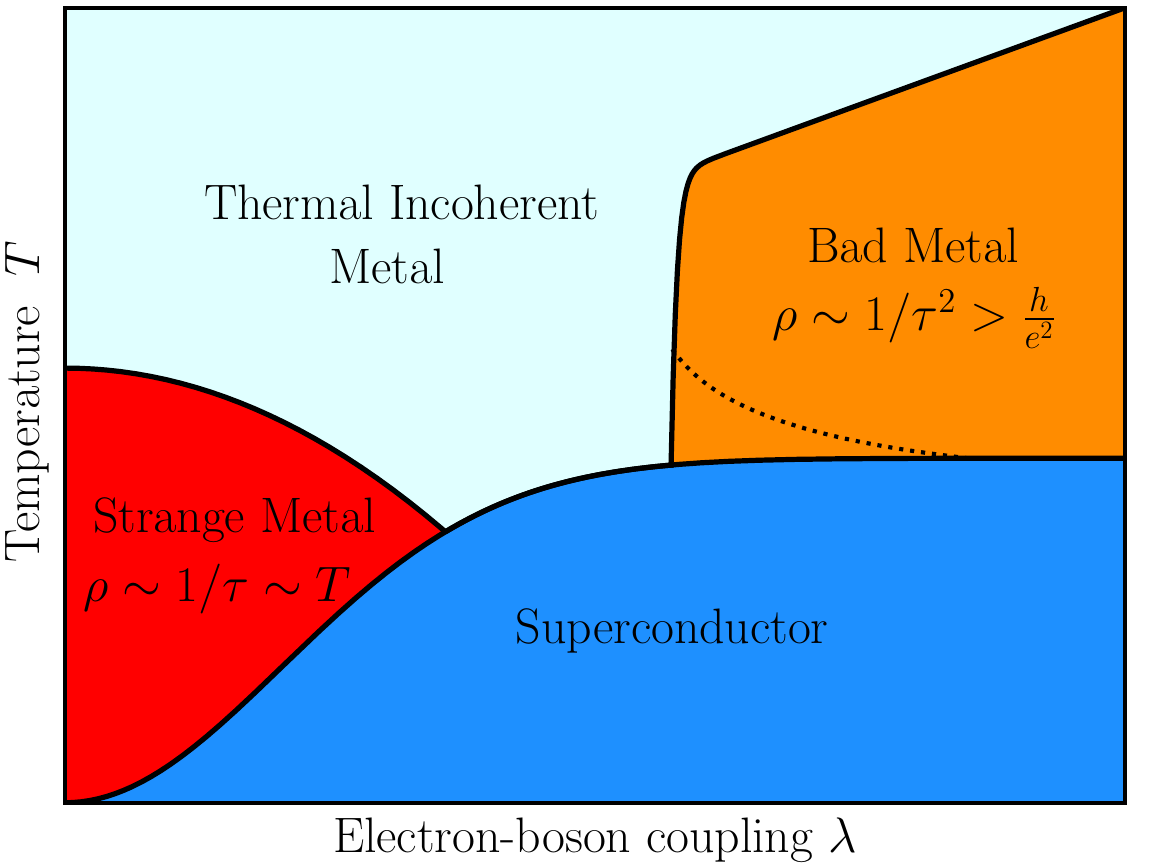}
\caption{Phase diagram for the (2+1)d Y-SYK model with spatially random electron-boson coupling tuned to the QCP. The KSS bound is violated above the dotted black line.}
\label{fig:PhaseDiagram}
\end{figure}

In this letter, we utilize the Y-SYK model to realize incoherent-metal phases. 
In particular, we will show that in the weak- and strong-coupling regimes, for finite and infinite bandwith, we obtain a phase diagram of strange and bad metals shown in Fig.~\ref{fig:PhaseDiagram}.

{\it The Model.---} The imaginary-time Lagrangian for the (2+1)d Yukawa-Sachdev-Ye-Kitaev (Y-SYK) model~\cite{EsterlisSchmalian2019, Wang2020} of $N$ electrons $\psi_{i\sigma}$ coupled to $N$ bosons $\phi_i$ is 
\begin{align} 
&\mathcal{L}=\psi_{i \sigma}^{\dagger}\left[\partial_\tau+\varepsilon(-i \nabla)-\mu\right] \psi_{i \sigma}+\frac{v_{i j}(\boldsymbol{r})}{\sqrt{N}} \psi_{i \sigma}^{\dagger} \psi_{j \sigma}  \label{eq:Lagrangian} \\
&+\frac{1}{2}\left[\left(\partial_\tau \phi_i\right)^2+c^2\left(\nabla \phi_i\right)^2+M_0^2 \phi_i^2\right] +\frac{g_{i j l}(\boldsymbol{r})}{N}  \phi_{l} \psi_{i \sigma}^{\dagger}  \psi_{j \sigma},\nonumber
\end{align}
where the flavor $i,j,l =1, \cdots, N$ and spin $\sigma=\uparrow \downarrow$ indices are summed over. The bare boson mass is denoted by $M_0$. Throughout, we will use Natural units: $\hbar=k_{B}=1$. The electron-boson interaction and potential are spatially disordered~\cite{Sachdev2024}, with coupling constants $g_{i j l}$ and $v_{ij}$ sampled from Gaussian orthogonal ensembles with zero mean and non-zero local variances, respectively:
\begin{align}
 \left\langle g_{i j l}(\bm{r}) g_{a b c}\left(\bm{r}^{\prime}\right)\right\rangle &= 2 g^2 \delta\left(\bm{r}-\bm{r}^{\prime}\right)\delta_{ia}\delta_{jb}\delta_{lc},  \label{eq:gVariance}\\ 
\langle v_{ij}(\bm{r}) v_{ab}(\bm{r}^{\prime})\rangle &= 2v^2\delta({\bm r}-{\bm r}')\delta_{ia}\delta_{jb}.
\label{eq:vVariance}
\end{align} 
In the large-$N$ limit, the normal-state saddle-point equations for the fermionic and bosonic self-energies of the disorder averaged~\cite{Dotsenko} model are given by $\Sigma(\tau)=(g^2\mathcal{D}(\tau) +v^2)\mathcal{G}(\tau)$ and $\Pi(\tau)=-2g^2\mathcal{G}(\tau)\mathcal{G}(-\tau)$, respectively. Due to the local nature of the variances in Eqs.~\eqref{eq:gVariance}-\eqref{eq:vVariance}, the momentum integrals for the fermionic and bosonic self-energies factorize (rather than the usual~\cite{Eliashberg1960, Eliashberg1961, Eliashberg1963, AGD, Karakozov1975, Marsiglio1988, MahanBook, Schmalian1996, Bennemann, Marsiglio2020} convolution), leading to manifestly momentum-independent self-energies. The effective fermionic and bosonic propagators $\mathcal{G}$ and $\mathcal{D}$ are~\cite{Gnezdilov2025, Esterlis2021}:
\begin{align}
\label{eq:Geff}
\mathcal{G}\left(i \omega_n\right) &= \int \frac{1}{i \omega_n-\xi_{\bm{k}}-\Sigma\left(i \omega_n\right)} \frac{d^2 k}{(2 \pi)^2},  \\  
\label{eq:Deff}
\mathcal{D}\left(i \Omega_m\right) &= \int_0^{\omega_D / c} \frac{1}{\Omega_m^2+c^2q^2+M_0^2-\Pi\left(i \Omega_m\right)} \frac{q d q}{2 \pi}.  
\end{align}
Here, $\omega_D$ and $c$ are analogous to the Debye frequency and the speed of sound, respectively. The ratio $\omega_D/c$ acts as a bosonic ultraviolet (UV) cutoff. In momentum space, the electronic dispersion relation is $\xi_{\bm{k}}=\varepsilon_{\bm{k}}-\mu;\;\varepsilon_{\bm{k}}\equiv k^2/(2m)\in[0,\Lambda]$, with bandwidth $\Lambda$ and chemical potential $\mu$. Here, we focus on the case of a symmetric bandwidth ($\mu=\Lambda/2$).

For weak coupling and low temperatures, the separation of scales $\Lambda/2  \gg  |i\omega_n-\Sigma(i\omega_n)|$ is satisfied, and the  Green's function in Eq.~\eqref{eq:Geff} is independent of the fermionic self-energy. The electronic and bosonic self-energies can then be evaluated exactly, and the bandwidth can be formally treated as infinite~\cite{Esterlis2021}. This approach has been widely used to study the strange-metal phase~\cite{Patel2023}, where the electronic self-energy in this regime takes the form of the marginal Fermi-liquid (MFL), where $\Sigma(i\omega)\sim-i\lambda\omega\log(1/|\omega|)$ for $T\ll\varepsilon_F$ and sufficiently weak coupling~\cite{Varma1989, Crisan1996}. Here, the dimensionless electron-boson coupling $\lambda\equiv\nu g^2/(4\pi c^2)$ and the electron density of states per spin is $\nu\equiv m/(2\pi)$ in 2d.

{\it The Bad Metal.---}
To understand the BM, we consider the case where $\Lambda$ is not the largest energy scale and the infinite bandwidth limit is inapplicable. When $\Sigma(i\omega_n) \gtrsim \Lambda$, the Green's function in Eq.~\eqref{eq:Geff} is not independent of the self-energy~\cite{Patel2018}. Expanding Eq.~\eqref{eq:Geff} to leading order in $\Lambda$, the effective electronic Green's function has an analogous form to the (0+1)d Y-SYK theory~\cite{EsterlisSchmalian2019, Wang2020}. In this regime, the electronic dispersion is inconsequential, which allows for a universal description for arbitrary bandstructure~\cite{Valentinis2026}. In the (0+1)d Y-SYK theory, at intermediate temperatures and coupling, a regime emerges where the electrons are highly incoherent and the bosons are soft and underdamped. In this regime, the thermal contribution to the fermionic self-energy is dominant and has an impurity-like form, while the bosons self-tune to the QCP for an arbitrary bare boson mass without the use of a constraint~\cite{EsterlisSchmalian2019, Wang2020, ClassenChubukov2021, Pan2021}. We find an analogous regime in the (2+1)d Y-SYK model, which enables us to realize a bad metal in the strong-coupling regime. 

For sufficiently strong coupling, the separation of scales $\Lambda/2 \ll |i\omega_n-\Sigma(i\omega_n)|$ is satisfied in the normal phase. Over a wide range of temperatures, electronic interactions provide the largest energy scale such that $\Lambda\tau\ll\hbar$, providing a realization of a bad metal. In this regime, the thermal contribution to the electronic self-energy is dominant $\Sigma(i\omega_n)\approx g^2T\mathcal{D}(0)\mathcal{G}(i\omega_n)$, giving rise to a fermionic self-energy in the form of an impurity potential:~$\Sigma(i\omega_n)\simeq-i\text{sgn}(\omega_n)\Omega_0/2$, where, in terms of the renormalized boson mass $M^2(T)\equiv M_0^2-\Pi(0)$, $\Omega_0/2\simeq\sqrt{2\lambda\Lambda T\log(\omega_D/M)}$. Upon evaluation of $\Pi(i\Omega_m)$, the bosons are found to be underdamped with a strongly renormalized boson mass $\log(\omega_D/M)\sim\Omega_0/T$ at low temperatures, so that $\Omega_0\sim\Lambda\lambda$ for $T\ll\Omega_0$. For high temperatures ($T\sim\Omega_0$), temperature-dependent corrections to the electronic scattering rate become relevant, and at ultra-high temperatures ($T\gg\Omega_0$), the renormalized bosonic mass approaches its bare value asymptotically $M(T)\sim M_0$, so that $\Omega_0\sim\sqrt{T}$.

\begin{figure}[t]
\centering\includegraphics[width=1\columnwidth]{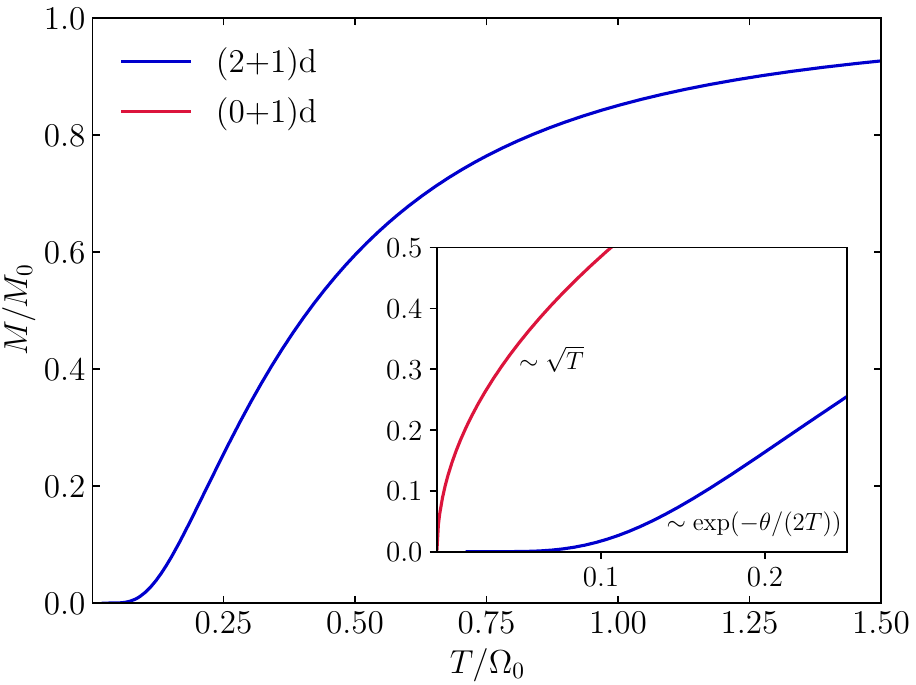}
\caption{Renormalized boson mass as a function of temperature $T/\Omega_0$ in the impurity regime for $M_0=\omega_D$. The inset shows the renormalized boson masses in (0+1)d and (2+1)d at low $T/\Omega_{0}$, where $\theta=\lambda\Lambda/(3\pi)^2$.} 
\label{fig:FBWBosonMass}
\end{figure}

Interestingly, the bosons are found to self-tune~\cite{Cichutek2024} to the quantum critical point (QCP) in the strong-coupling regime, as in the (0+1)d model~\cite{EsterlisSchmalian2019}. That is, $M(T)\rightarrow0$ as $T\rightarrow0$, for an arbitrary bare bosonic mass. Due to the momentum dependence of the bosonic dispersion, the bosonic propagator in (2+1)d has a logarithmic form~\cite{Gnezdilov2025} as opposed to the (0+1)d propagator~\cite{EsterlisSchmalian2019}, resulting in an exponential suppression of the renormalized boson mass at low $T$. A comparison between the (0+1)d and (2+1)d renormalized boson masses for the impurity regime is provided in Fig.~\ref{fig:FBWBosonMass}. For all figures in the main text, the energy scales are: $\omega_D=\varepsilon_F=\Lambda/2$. Another intriguing aspect of the self-tuning property of the (2+1)d Y-SYK model is that, due to the self-consistent evaluation of the bosonic mass, the fermionic scattering rate in the strong-coupling bad-metal phase is unaffected by spatially random mass disorder at low temperatures. There is an exact cancellation between the contribution of the potential disorder to the fermionic self-energy and the renormalized boson mass that appears in the electron-boson interaction~\cite{Supplement}.

{\it Non-Boltzmann Transport.---} In the Y-SYK model~\eqref{eq:Lagrangian}, the self-energy is independent of momentum and thus electromagnetic vertex corrections are absent~\cite{He2014, Ge2020, Raines2024}. The electromagnetic response is thus determined by a particle-hole bubble in terms of the dressed fermionic Green's function. The relevant energy scales for transport are temperature $T$, fermionic bandwidth $\Lambda$, and spectral width $\Gamma(\omega)\equiv |\Sigma_R^{\prime\prime}(\omega)|$. Here, $\Sigma_R(\omega)$ is derived solely from momentum and current-relaxing interactions. When the separation of scales $\Gamma(\omega)\ll\Lambda$ is satisfied, the excitations are sharply peaked and the fermionic spectral function is a Dirac delta function. The longitudinal DC electrical conductivity is thus given by the standard Drude formula $\sigma_{\text{Drude}}=e^2\nu v_F^2\tau_{\text{tr}}$, in terms of the Fermi-velocity, $v_F$, and transport lifetime, $\tau_{\text{tr}}$. This regime accurately describes conventional metals as well as unconventional metals, such as the strange metals, provided that the electronic scattering rate is much smaller than the electronic bandwidth. In the limit $T\ll\Gamma(\omega),\Lambda$, the single-particle lifetime $\tau \equiv 1/(2\Gamma)\approx 1/(2|\Sigma_R^{\prime\prime}(0))|$ is equivalent to the transport lifetime, defined by $\tau_{\text{tr}}\equiv2\pi \sigma/(e^2\Lambda)=\tau$. The resistivity is necessarily below the MIR bound, since $\rho\propto h^2(e^2 \Lambda \tau)^{-1}\ll h/e^2$. 

\begin{figure}[t]
\centering
\includegraphics[width=\linewidth]{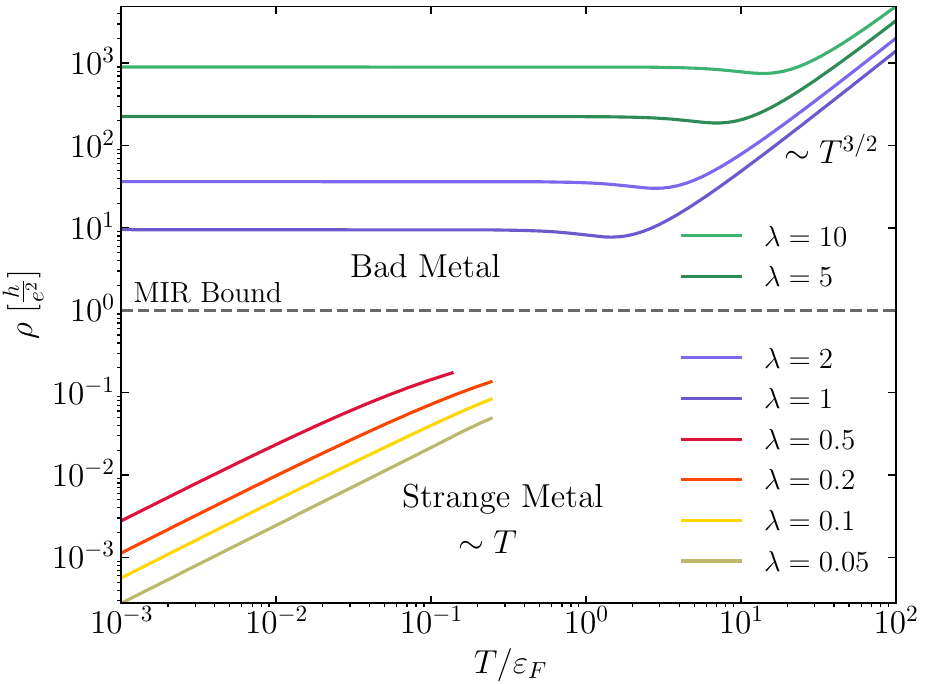}
\caption{(a) Resistivity (in units of $h/e^2$) as a function of temperature $T/\varepsilon_F$.} 
\label{fig:Rhofigure}
\end{figure}

In the limit $\Lambda\tau\sim \hbar$, sharp excitations no longer exist and the infinite-bandwidth limit is inapplicable. Beyond this regime, the separation of scales $T,\Lambda \ll \Gamma$, characterizes a strongly incoherent metal, where $A(\omega,\xi)\approx 2/\Gamma = 4\tau$ over the entire electronic band \footnote{Here, we assume $|\Sigma_R^{\prime\prime}(0)|\gg |\Sigma_R^{\prime}(0)|$.}. In this regime, the conductivity is given by~\footnote{The generalization to intermediate temperatures with a dynamical scattering rate is provided in the Supplemental Material~\cite{Supplement}.}
\begin{equation}
\sigma_{\BM} = 2\left(e\Lambda\tau/\pi\right)^2\hbar^{-3}.
\label{eq:non-boltzmann-transport}
\end{equation}
A similar relationship between $\tau$ and resistivity was noticed in Ref.~\cite{Parcollet1999}. The resistivity is necessarily greater than the MIR bound (since $\Lambda\tau\ll\hbar$), thus describing a bad metal. In this regime, the transport and single-particle lifetimes are related by: $\tau_{\text{tr}} = (4\Lambda/\pi) \tau^2$, which characterizes non-Boltzmann transport beyond the MIR limit~\cite{Supplement}. Using the formula above, the resistivity is $\rho_{\BM}=(\pi\Omega_0/(e\Lambda))^2/2$ in the strong-coupling regime for $T\ll\Omega_0$. In Fig.~\ref{fig:Rhofigure}, we plot the resistivity as a function of temperature for different coupling strengths. In the weak-coupling, low-$T$ regime we observe linear-$T$ resistivity consistent with strange-metal behavior, whereas for strong-coupling the resistivity surpasses the MIR bound. We refer to this regime as a bad metal, since $\rho>h/e^2$, although it does not display $T$-linear scaling. The quadratic dependence of the longitudinal conductivity $\sigma$ on $\tau$ leads to a manifest violation of Matthiessen's rule~\cite{ZimanBook, Maslov2017} in a bad metal, which has been previously discussed for strange metals~\cite{Liu2024}. 

At high temperatures, $\Lambda\ll\Gamma\ll T$, the thermal distribution function becomes broad and the conductivity and transport time relation crosses over to $\sigma_{\TIM}\sim e^2\Lambda^2 \tau/T$ and $\tau_{\text{tr}}\sim (\Lambda/T)\tau$, respectively. Here, TIM denotes the ``thermal incoherent metal'', which we ascribe to the ultra-high temperature regime. In this regime, we find $\rho_{\TIM}\sim T^{3/2}$, since $\Omega_0\sim\sqrt{T}$ at ultra-high temperatures.

{\it Violation of Transport Bounds.---} 
Studies of shear viscosity~\cite{Kadanoff1963, Luttinger1964, Hosoya1984, Taylor2010, Bradlyn2012, He2014} for interacting systems are extremely challenging, especially due to the inapplicability of perturbation theory~\cite{Jeon1995}. Renewed interest in shear viscosity is in part due to the KSS conjecture that the shear viscosity to entropy density ratio $\eta/s\geq\hbar/(4\pi k_{B})$. This conjecture has led to extensive searches for near-perfect fluid behavior~\cite{Schafer2009, Schafer2015, Zaanen2019} in ultra-cold Fermi gases~\cite{Thomas2010, Guo2011, Wlazlowski2015}, graphene~\cite{Muller2009}, quark-gluon plasmas~\cite{Demir2009}, and high-temperature superconductors~\cite{Rameau2014}. 

In Ref.~\cite{Eberlein2017} it was suggested that quantum critical theories strongly violate the KSS bound due to the presence of a well-defined Fermi surface. A study of perfect fluidity and incoherency in the context of the SYK model was performed by Ref.~\cite{Ge2020}, where it was found that at intermediate temperatures the KSS bound is violated. In this letter, we go beyond the work of Refs.~\cite{Eberlein2017, Ge2020} and consider the strong-coupling limit, which enables us to realize universal aspects of a bad metal.

For long-lived excitations, the shear viscosity is given by $\eta = \nu\varepsilon_F^2\tau$, where $\tau$ is the momentum-relaxing single-particle lifetime and we assumed $\Lambda/2=\varepsilon_F$. A quasi-particle description is based on the assumptions of long-lived ($\varepsilon_F\tau\gg\hbar$) and low-energy ($T\ll\varepsilon_F$) excitations. In Fermi-liquid theory~\cite{AGD}, the entropy density is $s\sim\nu T$ for $T\ll\varepsilon_F$. Consequently, $\eta/s$ is necessarily large: $\eta/s\sim(\varepsilon_F/T)\times(\varepsilon_F\tau)\gg\hbar/k_B$~\cite{Kovtun2003}. In general, even in superfluids~\cite{Abrikosov1959, Valls1974} or superconductors~\cite{Liao2019}, $\eta/s$ is expected~\cite{Kovtun2005} to be large due to the presence of long-lived quasiparticles. An important aspect of this work is that the saddle-point equations (and thus the transport expressions) are exact for all coupling strengths; that is, we do not perform diagrammatic perturbation theory~\cite{Luttinger1960}, and thus we can consider the strong-coupling and incoherent regime: $\lambda\gg1$ and $\Lambda\tau\ll\hbar$.

For strange metals, the entropy density has the form of a marginal Fermi liquid and is logarithmically enhanced at low temperatures, $s_{\SM}\sim\nu T\log(1/T)$~\cite{Crisan1996}. Although strange metals lack a conventional quasi-particle description, they satisfy the quasi-particle-like separation of scales, $\Lambda\tau\gg\hbar$ (for temperatures $T\ll\Lambda$), implying that the shear viscosity is proportional to the scattering rate, as in the case for long-lived excitations. Thus, the ratio $\eta/s$ is quite large for temperatures $T\ll\Lambda$~\cite{Supplement}. 

In the regime $\Lambda\tau\lesssim\hbar$, the Drude-like formula for shear viscosity is inapplicable. For a bad metal obeying $\Lambda\tau\ll\hbar$, we find, in analogy to Eq.~\eqref{eq:non-boltzmann-transport}, that shear viscosity is described by a non-Boltzmann formula:
\begin{equation}
\eta_{\BM}=4\nu\Lambda^3\tau^2/(3\pi \hbar).
\end{equation}
In the bad-metal regime, the entropy density is $s_{\text{BM}}=s_{\text{Int}}+s_{B,0}$, where the interaction and free-boson contributions are $s_{\text{Int}}\sim\nu \Lambda T/\Omega_0$ and $s_{B,0}\sim\nu\varepsilon_F(T/\omega_D)^2$ respectively. In the strong-coupling limit, $\Lambda\tau\ll\hbar$ is satisfied and the bosons provide the dominant contribution to the entropy density. Physically, this result arises because, in proximity to the QCP, the renormalized boson mass is exponentially suppressed, so that $M(T)\ll T$, whereas for the fermions $\Omega_0 \gg T$ for intermediate temperatures. In this sense, the bosons are `hot' and the electrons are `cold' at intermediate temperatures~\cite{EsterlisSchmalian2019}, which leads to the bosons providing the dominant contribution to the entropy density in the strong-coupling limit.

\begin{figure}[t]
\centering
\includegraphics[width=\linewidth]{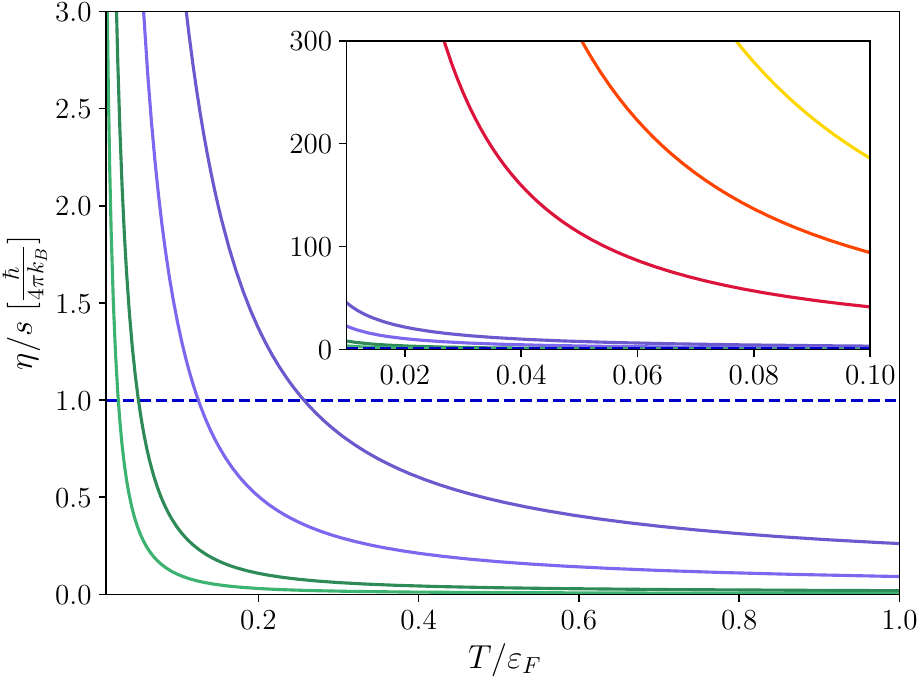}
\caption{Shear viscosity to entropy density ratio $\eta/s$ (in units of $\hbar/(4\pi k_B)$) for $T>0.01\varepsilon_F$. Subplot shows $\eta/s$ in the weak-coupling regime. Dashed blue line denotes the KSS bound: $\hbar/(4\pi k_B)$. The values of $\lambda$ are given in the legend of Fig.~\ref{fig:Rhofigure}.}
\label{fig:Etafigure}
\end{figure}

\begin{figure}[t]
\centering\includegraphics[width=1\columnwidth]{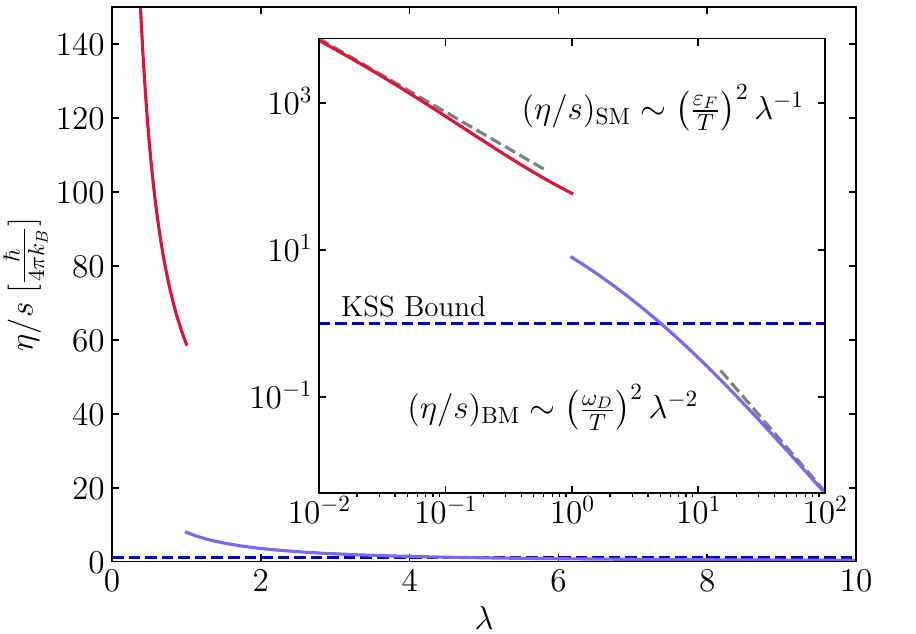}
\caption{Shear viscosity to entropy density ratio $\eta/s$ (in units of $\hbar/(4\pi k_B)$) as a function of dimensionless coupling, $\lambda$, evaluated at $T=0.05\varepsilon_F$. Red and blue curves denote $\eta/s$ for the infinite and finite bandwidth limit solutions respectively. Inset displays $\eta/s$ on log-log scales. Dashed blue and gray lines denotes the KSS bound: $\hbar/(4\pi k_B)$ and asymptotic values respectively.}
\label{fig:EtaS-coupling}
\end{figure}

In the bad-metal regime, for strong-coupling ($\lambda\gg1$) and low temperatures ($T\ll\omega_D$), we find $(\eta/s)_{\BM}\sim (\omega_D/T)^{2}/\lambda^2$. At intermediate temperatures, the KSS bound is strongly violated, as shown in Fig~\ref{fig:Etafigure}. A small $\eta/s$ relative to the KSS bound appears to be a generic feature of bad metals, since the shear viscosity is very small: $\eta\ll\nu\Lambda$. In Figs.~\ref{fig:Etafigure} and~\ref{fig:EtaS-coupling} we plot $\eta/s$ as functions of temperature and coupling, respectively.

There has been discussion~\cite{Schafer2015} about whether momentum diffusion is best encapsulated in the ratio $\eta/s$ or the diffusion constant $D$. Indeed, in certain cases~\cite{Herzog2006} the latter is zero in the strong-coupling limit, whereas the ratio $\eta/s$ remains finite. Moreover, as noted in Ref.~\cite{Zaanen2019}, there are possible distinctions in the definitions of viscosity via the Kubo formula and diffusivity~\cite{Son2007}. Hartnoll~\cite{Hartnoll2015} has suggested that transport in incoherent metals is described by diffusion of charge and energy, not by momentum diffusion, and he proposed the general bound for the diffusivity $D\gtrsim v_F^2/T$. We computed the charge diffusivity for the strongly incoherent metal and found that $D\sim v_F^2\tau$ in the strong-coupling regime (both at low and ultra-high temperatures), which agrees with the standard result~\cite{Hartman2017} for conventional metals up to a numerical coefficient. For the bad and thermal incoherent metal regime, the diffusivity bound is violated for $T\lesssim\Omega_0$. Violation of the diffusivity bound was also found in other contexts~\cite{Pakhira2015, Werman2016}.

As shown in Fig.~\ref{fig:PhaseDiagram}, we find that strange- and bad-metal phases exist in the weak and strong-coupling regimes, respectively. With increasing temperature and coupling, the strange-metal regime becomes incoherent as the solutions depart from the MFL form, setting the upper boundary for the region. For sufficiently strong coupling, the intermediate temperature solutions describe a bad metal with strong impurity physics. Across all nonzero values of coupling, the high-temperature regime is incoherent and dominated by thermal effects. At low temperatures, superconductivity exists with a critical temperature monotonically increasing with electron-boson coupling~\cite{Gnezdilov2025}, before saturating to a constant~\cite{Supplement}.

{\it Conclusions.---} This paper has provided the first step in realizing a microscopic model that exhibits numerous universal aspects of bad-metal behavior in the electron-boson framework. By using a finite electronic bandwidth, we showed that the (2+1)d Y-SYK model also exhibits an impurity regime (as in the (0+1)d model), where the boson mass self tunes to quantum criticality and the fermionic self-energy provides the largest energy scale. In this regime, we showed that the resistivity exceeds the MIR bound, the shear viscosity to the entropy density can exceed the KSS bound, and the transport formulae are described by a non-Boltzmann formula that exhibits the incoherent nature of the system. Our analysis also shows how the strange-metal phase crosses over to the bad-metal phase, providing an important step toward understanding the normal phases of unconventional superconductors.

{\it Acknowledgment.---} This research was supported by funding provided by Dartmouth College.

\bibliography{References}

\end{document}


\title{Supplemental Material: Universal Theory of Incoherent Metals}

\author{Aaron Kleger}
\email{Aaron.R.Kleger.GR@dartmouth.edu}
\author{Nikolay Gnezdilov}
\email{Nikolay.Gnezdilov@dartmouth.edu}
\author{Rufus Boyack}
\email{Rufus.Boyack@dartmouth.edu}

\affiliation{Department of Physics and Astronomy, Dartmouth College, Hanover, New Hampshire 03755, USA}

\maketitle

\tableofcontents

\section{The model}

\subsection{Normal-state Green's functions}

We consider the Yukawa-Sachdev-Ye-Kitaev (Y-SYK) model~\cite{EsterlisSchmalian2019, Wang2020} in two spatial dimensions, consisting of $N$ electrons $\psi_{i\sigma}$ coupled to $N$ bosons $\phi_i$ in the continuous limit; $\sigma=\uparrow \downarrow$ denotes spin, $i=1, \cdots, N$ labels the fermion and boson flavor. The Hamiltonian reads $H=H_\psi+H_\phi+V_{\psi \phi}$, where
\begin{subequations}
\begin{align}
H_\psi & =\sum_{i=1}^N \sum_{\sigma=\uparrow \downarrow} \int \psi_{i \sigma}^{\dagger}(\bm{k}) \xi_{\bm{k}}\psi_{i \sigma}(\bm{k}) \frac{d^2 k}{(2 \pi)^2}, \\
H_\phi & =\frac{1}{2} \sum_{i=1}^N \int  \phi_i(\bm{q}) \left(c^2q^2+M_0^2\right) \phi_i(-\bm{q}) \frac{d^2 q}{(2 \pi)^2}, \\
V_{\psi \phi} & =\frac{1}{N} \sum_{i j l=1}^N \sum_{\sigma=\uparrow \downarrow} \int g_{i j l}(\mathbf{r}) \psi_{i \sigma}^{\dagger}(\mathbf{r}) \psi_{j \sigma}(\mathbf{r}) \phi_l(\mathbf{r}) d^2 r.
\end{align}
\end{subequations}
The electronic dispersion relation is given by $\xi_{\bm{k}}=k^2/(2m)-\mu;\;k^2/(2m)\in[0,\Lambda]$, with bandwidth $\Lambda$ and chemical potential $\mu$. The bare mass of the boson is denoted by $M_0$. Throughout, we will use Natural units: $\hbar=k_{B}=1$. The electron-boson interaction is spatially disordered~\cite{Sachdev2024}, with coupling constants $g_{i j l}$ sampled from a Gaussian orthogonal ensemble with zero mean and finite local variance $\left\langle g_{i j l}(\mathbf{r}) g_{a b c}\left(\mathbf{r}^{\prime}\right)\right\rangle = 2 g^2 \delta\left(\mathbf{r}-\mathbf{r}^{\prime}\right)\delta_{ia}\delta_{jb}\delta_{lc}$. 

In the large-$N$ limit, the normal-state saddle-point equations for the fermionic and bosonic self-energies are~\cite{Gnezdilov2025}:
\begin{subequations}
\begin{align}
\Sigma\left(i \omega_n\right) & =g^2 T \sum_{\omega_{n^{\prime}}} \mathcal{D}\left(i \omega_n-i \omega_{n^{\prime}}\right) \mathcal{G}\left(i \omega_{n^{\prime}}\right), \label{eq:Sigma_saddlept} \\ 
\Pi\left(i \Omega_m\right) & =-2 g^2 T \sum_{\omega_n} \mathcal{G}\left(i \omega_n\right) \mathcal{G}\left(i \omega_n+i \Omega_m\right),
\label{eq:Pi-saddlept}
\end{align}
\end{subequations}
where $\omega_n=(2 n+1) \pi T$ and $\Omega_m=2 m \pi T$ are fermionic and bosonic Matsubara frequencies, respectively, at temperature $T$. To decouple the interactions, $\Sigma$ and $\Pi$ are introduced as Lagrange multipliers that are non-local in time. Due to the local variance of the electron-boson coupling, the momentum integrals for the fermionic and bosonic self-energies factorize (rather than the usual~\cite{AGD} convolution), leading to manifestly momentum-independent self-energies.

The effective fermionic and bosonic propagators $\mathcal{G}$ and $\mathcal{D}$ are defined by
\begin{align} 
    \mathcal{G}\left(i \omega_n\right) & \equiv \int G\left(i \omega_n, \bm{k}\right) \frac{d^2 k}{(2 \pi)^2} \nonumber \\  
& =\int \frac{1}{i \omega_n-\xi_{\bm{k}}-\Sigma\left(i \omega_n\right)} \frac{d^2 k}{(2 \pi)^2}. \label{eq:G-eff} \\  
\mathcal{D}\left(i \Omega_m\right) & \equiv \int D\left(i \Omega_m, \bm{q}\right) \frac{d^2 q}{(2 \pi)^2} \nonumber \\
& =\int_0^{\omega_D / c} \frac{1}{\Omega_m^2+c^2q^2+M_0^2-\Pi\left(i \Omega_m\right)} \frac{q d q}{2 \pi}.
\label{eq:Deff}
\end{align}
Here, $\omega_D$ and $c$ are analogous to the Debye frequency and the speed of sound, respectively. The ratio $\omega_D/c$ acts as a bosonic ultraviolet (UV) cutoff; it is also equivalent to a Pauli-Villars regulator~\cite{Gnezdilov2025, Esterlis2021}. Performing the momentum integration in Eq.~\eqref{eq:Deff}, the result obtained is $\mathcal{D}\left(i \Omega_m\right)=L\left(i \Omega_m\right) /\left(4 \pi c^2\right)$, where
\begin{equation}
L\left(i \Omega_m\right)=\log \left[1+\frac{\omega_D^2}{\Omega_m^2 +M_0^2-\Pi(i\Omega_m)}\right].
\label{eq:BosonL}
\end{equation}
Using a symmetric bandwidth ($\mu=\Lambda/2$) and assuming $\text{Re}\Sigma(i\omega_n)=0$ ensures that the real part of the effective fermionic Green's function is zero. In conventional electron-phonon systems, this assumption is equivalent to the even part of the Matsubara-axis fermionic self energy being zero~\cite{Bennemann}. Let $\nu=m /(2 \pi)$ be the electron density of states per spin in two-dimensions and $m$ the electron mass. After performing the $k$-integration in Eq.~\eqref{eq:G-eff}, we obtain:
\begin{equation}
\mathcal{G}(i\omega_n)=-2i\nu\ \text{arctan}\left(\frac{\Lambda/2}{\omega_n+i\Sigma(i\omega_{n})}\right).
\label{eq:G_Eff_ArbitraryBW}
\end{equation}

\subsection{Green's functions in the infinite-bandwidth limit}

Let $x=(\Lambda/2)/(\omega_n+i\Sigma(i\omega_n))$. For $|x|\gg1$, $\text{arctan}(x)\sim \frac{\pi}{2}\text{sgn}(x)$. In the infinite bandwidth limit, the effective Green's function depends only on frequency and is independent of the fermionic self-energy. Thus, when the separation of scales $\Lambda/2  \gg  |i\omega_n - \Sigma(i\omega_n)|$ is satisfied, the  effective electronic Green's function in Eq.~\eqref{eq:G-eff} can be evaluated by extending the momentum integration limits to positive and negative infinity. In this limit,  $\mathcal{G}\left(i \omega_n\right)=-i \pi \nu \operatorname{sgn}\left(\omega_n\right)$.

At nonzero temperatures, a nonzero boson mass is necessary to ensure that the thermal contribution (the resonant term with $\omega_{n^\prime}=\omega_n$ in Eq.~\eqref{eq:Sigma_saddlept}) to the fermionic self-energy is finite. To ensure that the renormalized boson mass, defined by $M^{2}(T)\equiv M_0^2-\Pi(0)$, vanishes at $T=0$ and is nonzero for $T>0$, we introduce a constraint in the action~\cite{Esterlis2021}. The renormalized boson mass can be tuned to quantum criticality by replacing the bare bosonic mass with a Lagrange multiplier $M_0^2\rightarrow\lambda(\tau,\bm{r})$, and adding an additional term to the action~\cite{Esterlis2021}:
\begin{equation}
S_\gamma = -N\int  \frac{i\lambda(\tau,\bm{r})}{2\gamma}d\tau d^2r.
\end{equation}
The constraint yields the additional saddle-point equation
\begin{equation}
\frac{1}{\gamma} = T\sum_{\Omega_m}\mathcal{D}(i\Omega_m),
\label{eq:Constraint_saddlept}
\end{equation}
and $i\lambda(\tau,\bm{r})=M_0^2$ at the large-$N$ saddle-point solution. The renormalized boson mass is found by equating the right-hand side of Eq.~\eqref{eq:Constraint_saddlept} at zero and nonzero temperature (as in Eq.~\eqref{eq:Boson-constraint}), then solving for $M^2(T)$ (see Sec.~\ref{sec:RenormalizedBosonMass} for further details). For a specific value of $\gamma$, the renormalized boson mass is zero at zero temperature and nonzero at nonzero temperatures. The constraint in the action gives the following additional term in the free energy~\footnote{Volume is not explicitly introduced in this manuscript (akin to unit lattice spacing): $[k_F]=1$. Thus, electronic bandwidth and the inverse of the electron mass are in units of energy: $[\Lambda]=[m^{-1}]$.}:
\begin{equation} 
F_\gamma = -N\frac{M_0^2(T)}{2\gamma}.
\label{eq:FreeEnergyConstraint}
\end{equation}
The dynamical part of the bosonic self-energy is determined to be $\Pi\left(i \Omega_m\right)-\Pi(0)=$ $-2 \pi \nu^2 g^2\left|\Omega_m\right|$, where $\Pi(0) = M_0^2-M^2$ \cite{Esterlis2021,Gnezdilov2025}. Using the temperature independence of $\Pi(0)$, the constraint contribution to the free energy appearing in Eq.~\eqref{eq:FreeEnergyConstraint} may be more conveniently re-expressed as $F=-NM^2(T)/2\gamma$, in terms of the renormalized boson mass. 

As in Ref.~\cite{Gnezdilov2025}, we define the dimensionless electron-boson coupling constant $\lambda=\nu g^2 /\left(4 \pi c^2\right)$, where it was assumed that the Debye and Fermi momenta are of the same order $k_D\sim k_F$. In the infinite-bandwidth limit, $L(i\Omega_m)$ is 
\begin{equation}
L\left(i \Omega_m\right)=\log \left[1+\frac{1}{\left(\Omega_m/\omega_D\right)^2+\left(M/\omega_D\right)^2+2 \pi \lambda \left|\Omega_m\right|/\varepsilon_F}\right].
\label{eq:logarithm}
\end{equation}

\subsection{Green's functions in the finite-bandwidth limit}

For $|x|\ll1$, $\text{arctan}(x)\sim x$. In this regime, the effective electronic Green's function has an analogous form as in the (0+1)-dimensional theory~\cite{EsterlisSchmalian2019}. For this case, the electronic bandwidth $\Lambda$ is not the largest energy scale, and so the separation of scales $\Lambda/2  \gg  |i\omega_n - \Sigma(i\omega_n)|$ is not satisfied. When $\Sigma(i\omega_n)\sim \mathcal{O}(\Lambda)$ or greater, the effective electronic Green's function cannot be evaluated by performing momentum-space integration over an infinite-sized Fermi surface. In particular, the momentum-integrated electronic Green's function is not independent of its self-energy~\cite{Patel2018}. 
Expanding Eq.~\eqref{eq:G_Eff_ArbitraryBW} to leading order in $\Lambda$, the effective electronic Green's function is
\begin{equation}
\mathcal{G}(i\omega_n)=\frac{\nu\Lambda}{i\omega_n-\Sigma(i\omega_n)}.
\label{eq:G_CL}
\end{equation}
Furthermore, when it is valid to expand the logarithm in Eq.~\eqref{eq:logarithm} to lowest order in $\omega_D$, the effective bosonic Green's function becomes
\begin{equation}\label{eq:D-eff-expanded}
\mathcal{D}(i\Omega_m)=\frac{\omega_D^2}{4\pi c^2}\frac{1}{\Omega_m^2+M_0^2-\Pi(i\Omega_m)}.
\end{equation}
Thus, the normal-state saddle-point equations of the (0+1)-dimensional Y-SYK theory~\cite{EsterlisSchmalian2019, Wang2020} with $\mu=0$ are recovered. Although the saddle-point equations reduce to those of the lower dimensional theory, the full electronic Green's function, $G(i\omega_n,\bm{k})$, relevant for transport and thermodynamics, retains its 2-dimensional spatial structure.

The lower-dimensional theory admits~\cite{Esterlis2021} a quantum critical SYK-NFL fixed point at low temperatures for weak coupling, and an impurity-like NFL fixed point at intermediate temperatures for strong coupling. In Sec.~\ref{sec:ImpurityRegime} we find an analogous impurity regime as a solution of the (2+1)d Y-SYK saddle-point equations. Our strong-coupling analysis of the normal-state shear viscosity to entropy density ratio near superconducting $T_c$ is based on the impurity-like NFL solution in the finite-bandwidth limit.

\section{Renormalized boson mass}
\label{sec:RenormalizedBosonMass}

\subsection{Renormalized boson mass in the infinite-bandwidth limit}
\label{sec:BosonMassIBWL}

Equating the right-hand side of Eq.~\eqref{eq:Constraint_saddlept} at zero and nonzero temperatures:
\begin{equation} \label{eq:Boson-constraint}
\frac{1}{2\pi}\int_{-\infty}^{\infty} \left. L(i\Omega)  d\Omega \right|_{M=0}= T\sum_{\Omega_m}L(i\Omega_m)|_{M=M(T)}, 
\end{equation}
we can solve for the renormalized boson mass $M(T)$ at nonzero temperatures. Tuning the mass to quantum criticality, we set the zero-temperature renormalized bosonic mass to zero and divide by $\omega_{D}$ to obtain
\begin{equation}
\frac{1}{2\pi}\bigintssss_{-\infty}^{\infty} \log \left[1+\frac{1}{\left(\Omega/\omega_D\right)^2+2 \pi \lambda \left|\Omega\right|/\varepsilon_F}\right]\frac{d\Omega}{\omega_{D}} 
= \frac{T}{\omega_{D}}\sum_{\Omega_m}\log \left[1+\frac{1}{\left(\Omega_m /\omega_D\right)^2 +\left(M/\omega_D\right)^2+2 \pi \lambda \left|\Omega_m\right|/\varepsilon_F}\right].
\label{eq:BosonMassSelf-consistentEq}
\end{equation}
In the limit where $\lambda_E\equiv(\omega_D/\varepsilon_F)\lambda\ll1$, the dynamical part of the bosonic self-energy can be ignored, and the left-hand side of Eq.~\eqref{eq:BosonMassSelf-consistentEq} is
\begin{align}
\frac{1}{2 \pi}\int_{-\infty}^{\infty}\log \left[1+\frac{1}{\left(\Omega / \omega_D\right)^2+2 \pi \lambda|\Omega| / \varepsilon_F}\right] \frac{d \Omega}{\omega_{D}} 
& =\frac{1}{\pi}\int_0^{\infty} \log \left(1+\frac{1}{y^2+2 \pi \lambda y \omega_D / \varepsilon_F}\right) dy \nonumber\\ 
& \approx \frac{1}{\pi} \int_0^{\infty} \log \left(1+1/y^2\right) dy \nonumber \\ 
& =1.
\label{eq:LHS-NoPi}
\end{align}
Let $\alpha \equiv 2 \pi \lambda \omega_D / \varepsilon_F$. For arbitrary $\lambda_E$, the zero-temperature integral is evaluated as follows:
\begin{align}
\varepsilon(\alpha)\equiv\frac{1}{\pi} \bigintssss_0^{\infty} \log \left(1+\frac{1}{y^2+\alpha y}\right) d y
&=\frac{1}{\pi}\bigintssss_{0}^{1}\left\{\bigintssss_{0}^{\infty}\left(\frac{1}{y^2+\alpha y+s}\right)dy\right\} ds \nonumber\\
&=\frac{1}{\pi}\bigintssss_{0}^{1}\left[\frac{2}{\sqrt{4s-\alpha^2}}\left.\arctan\left(\frac{2y+\alpha}{\sqrt{4s-\alpha^2}}\right)\right|_{y=0}^{y=\infty}\right]ds \nonumber\\
&=\frac{1}{\pi}
\left\{\begin{array}{r} \alpha \log (\alpha) - \sqrt{\alpha^2-4}\ \textrm{arcoth}\left(\dfrac{\alpha}{\sqrt{\alpha^2-4}}\right), \quad \alpha\geq2 \\ \alpha \log (\alpha)+\sqrt{4-\alpha^2}\left(\pi / 2-\arctan \left(\dfrac{\alpha}{\sqrt{4-\alpha^2}}\right)\right), \quad 0 \leq \alpha \leq 2.\end{array}\right.
\label{eq:Zero-T-arb}
\end{align}

\begin{figure}[t]
\centering\includegraphics[width=.5\columnwidth]{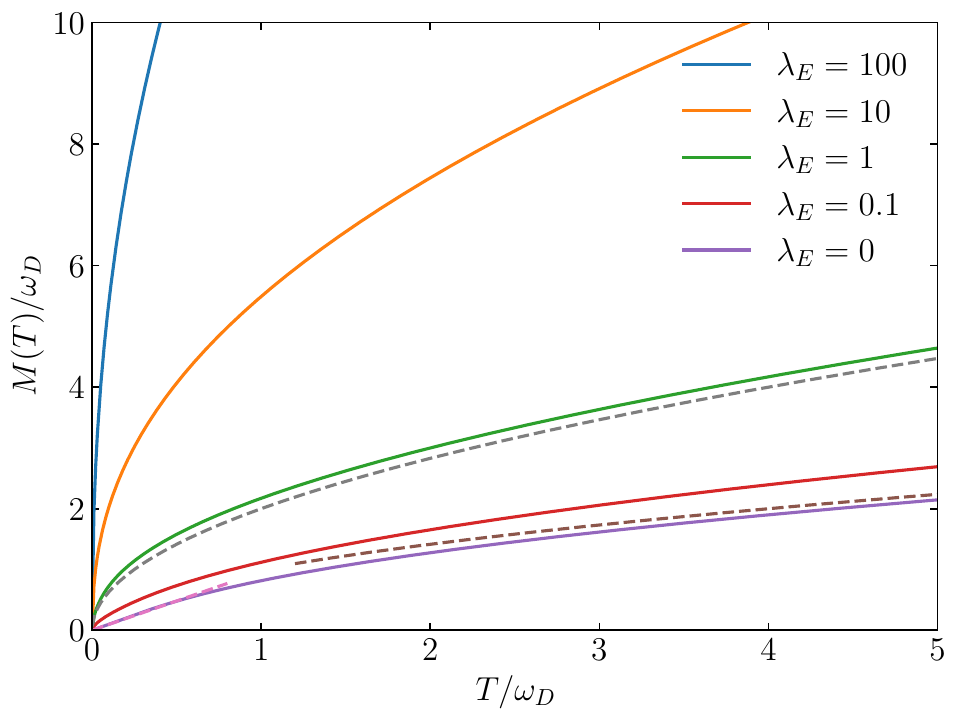}
\caption{Renormalized boson mass (in  units of $\omega_D$) in the infinite bandwidth limit as a function of temperature, $T/\omega_D$, for various values of dimensionless coupling $\lambda_E\equiv(\omega_D/\varepsilon_{F})\lambda$. Dashed pink, brown, and gray curves correspond to $\textrm{arcosh}\left(3/2\right)T/\omega_D$, $\sqrt{T/\omega_D}$, and $2\sqrt{T/\omega_D}$ respectively. }
\label{fig:BosonMass}
\end{figure}
The numerical solution of Eq.~\eqref{eq:BosonMassSelf-consistentEq} for $M(T)$ is shown in Fig.~\ref{fig:BosonMass}. The asymptotic behavior of $M(T)$ for high temperatures is determined by the thermal contribution (the leading term, $m=0$) in the summation in Eq.~\eqref{eq:BosonMassSelf-consistentEq}. Let $\bar{m}\equiv M/\omega_D,\;t\equiv T/\omega_D$. Considering only the $m=0$ term in the summation, the self-consistent equation for the renormalized boson mass takes the form $\varepsilon(\alpha)=t \log(1+1/\bar{m}^2)$. Solving for the renormalized boson mass, we obtain $\bar{m}^2(t)=n_B(\varepsilon(\alpha))$, where $n_B(x)\equiv1/(e^{x/t}-1)$. At high temperatures ($t \gg \varepsilon$), $\bar{m}(t)\sim\sqrt{t/\varepsilon(\alpha)}$ as found in Ref.~\cite{Esterlis2021}. For $\lambda_E=0$, the asymptotic behavior is $\bar{m}(t)\sim\sqrt{t}$, as indicated by the dashed brown curve in Fig.~\ref{fig:BosonMass}. In the asymptotically strong coupling limit, $\varepsilon(\alpha)\sim2\log\alpha/(\pi\alpha)$; therefore, the renormalized boson mass is $\bar{m}(t)\sim\pi\sqrt{t \lambda_E /\log(2\pi\lambda_E)}$ in the limit of asymptotically strong coupling and high temperature.

The low-temperature behavior in Fig.~\ref{fig:BosonMass} can be recovered by performing the Matsubara summation in Eq.~\eqref{eq:BosonMassSelf-consistentEq}. For small $\lambda_E$, the series in Eq.~\eqref{eq:BosonMassSelf-consistentEq} may be expressed as 
\begin{equation}
t\sum_{m=-\infty}^{\infty} \log \left(1+\frac{1}{4 \pi^2 m^2 t^2+\bar{m}^2+4 \pi^2 \lambda|m| t \omega_D / \varepsilon_F}\right) \approx  t\left[2\sum_{m=0}^{\infty} \log \left(1+\frac{1}{4 \pi^2 m^2 t^2+\bar{m}^2}\right) - \log \left(1+\frac{1}{\bar{m}^2}\right)\right].
\label{eq:RHS-NoPi}
\end{equation}
Equating Eqs.~\eqref{eq:LHS-NoPi} and~\eqref{eq:RHS-NoPi}, dividing by $t$, and then taking the exponential of both sides, we obtain
\begin{align}\nonumber
\exp \left(\frac{1}{t}\right)&=\exp \left\{2\sum_{m=0}^{\infty} \log \left(1+\frac{1 }{4 \pi^2 m^2t^2+\bar{m}^2}\right)-\log\left(1+\frac{1}{\bar{m}^2}\right)\right\}\\ \nonumber
&=\left(1+\frac{1}{\bar{m}^2}\right)^{-1}\prod_{m=0}^{\infty}\left(1+\frac{1}{4 \pi^2 m^2 t^2+\bar{m}^2}\right)^2\\
&=\left(1+\frac{1}{\bar{m}^2}\right)^{-1}\left(\frac{\sqrt{1+\bar{m}^2 } \sinh\left(\sqrt{1+\bar{m}^2}/(2 t) \right)}{\bar{m} \sinh \left(\bar{m}/(2 t)\right)}\right)^2
=\frac{\text{cosh}\left(\sqrt{1+\bar{m}^2}/t\right)-1}{\text{cosh}\left(\bar{m}/t\right)-1}.
\end{align}
In the third equality, we used the Weierstrass factorization theorem~\cite{WhittakerWatsonBook}. This transcendental equation for $\bar{m}\equiv M/\omega_D$ provides the asymptotic behavior at low-temperature displayed in the purple curve of Fig.~\ref{fig:BosonMass}. In the low-temperature limit, we obtain a function that is linear in $T$: $M/T \sim \text{arcosh}\left(3/2\right)$, indicated by the dashed pink curve in Fig.~\ref{fig:BosonMass}. 

As in Ref.~\cite{Esterlis2021}, the constraint equation can be expressed as a closed-form transcendental equation. The constraint is given in Eq.~\eqref{eq:BosonMassSelf-consistentEq}. The infinite summation appearing in this expression can be evaluated as follows
\begin{align}
S &= 2\sum_{m=0}^{\infty}\log\left[1+\frac{1}{(\Omega_{m}/\omega_{D})^{2}+(M/\omega_{D})^{2}+2\pi\lambda|\Omega_{m}|/\varepsilon_{F}}\right] \nonumber\\ 
&= 2\sum_{m=0}^{\infty}\int_{0}^{1}\frac{1}{(\Omega_{m}/\omega_{D})^{2}+(M/\omega_{D})^{2}+2\pi\lambda|\Omega_{m}|/\varepsilon_{F}+y}dy \nonumber\\ 
&= 2\int_{0}^{1}\sum_{m=0}^{\infty}\frac{1}{4\pi^{2}m^{2}t^{2}+\bar{m}^{2}+4\pi^{2}\lambda_{E}mt+y}dy \nonumber\\ 
&= 2\int_{0}^{1}\frac{1}{\sqrt{b^{2}-4ac}}\left[\psi\left(\frac{b+\sqrt{b^{2}-4ac}}{2a}\right)-\psi\left(\frac{b-\sqrt{b^{2}-4ac}}{2a}\right)\right] dy.
\end{align}
Here we define $a=4\pi^{2}t^{2},b=4\pi^{2}\lambda_{E}t,c=y+\bar{m}^{2}$ and use the digamma function~\cite{WhittakerWatsonBook} to perform the frequency summation.  Next, evaluate the integral using the definition of the digamma function, $\psi(z)=d\log\Gamma(z)/dz$, to obtain 
\begin{equation}
S=2\log\left[\dfrac{\Gamma\left(\dfrac{b-\sqrt{b^{2}-4a\bar{m}^{2}}}{2a}\right)}{\Gamma\left(\dfrac{b-\sqrt{b^{2}-4a(1+\bar{m}^{2})}}{2a}\right)}\dfrac{\Gamma\left(\dfrac{b+\sqrt{b^{2}-4a\bar{m}^{2}}}{2a}\right)}{\Gamma\left(\dfrac{b+\sqrt{b^{2}-4a(1+\bar{m}^{2})}}{2a}\right)}\right].
\end{equation}
Therefore, the transcendental equation for the bosonic mass $\bar{m}\equiv M/\omega_{D}$ can be expressed as 
\begin{equation}
\varepsilon(\alpha)	= 2t\log\left[\frac{\Gamma\left(\dfrac{\lambda_{E}}{2t}\left(1-\sqrt{1-\bar{m}^{2}/(\pi\lambda_{E})^{2}}\right)\right)}{\Gamma\left(\dfrac{\lambda_{E}}{2t}\left(1-\sqrt{1-(1+\bar{m}^{2})/(\pi\lambda_{E})^{2}}\right)\right)}\dfrac{\Gamma\left(\dfrac{\lambda_{E}}{2t}\left(1+\sqrt{1-\bar{m}^{2}/(\pi\lambda_{E})^{2}}\right)\right)}{\Gamma\left(\dfrac{\lambda_{E}}{2t}\left(1+\sqrt{1-(1+\bar{m}^{2})/(\pi\lambda_{E})^{2}}\right)\right)}\right] -t \log\left(1+\frac{1}{\bar{m}^{2}}\right).
\end{equation}

\subsection{Renormalized boson mass in the finite-bandwidth limit}
\label{sec:ImpurityRegime}

\subsubsection*{Intermediate temperatures}
Here, we will assume the separation of scales $|i\omega_n-\Sigma(i\omega_n)| \gg \Lambda/2$, and utilize the effective Green's function in Eq.~\eqref{eq:G_CL}. By neglecting the dynamical part of the bosonic self-energy, $\Pi(i\Omega_{m})-\Pi(0)$, and assuming $T \gg M(T)$, the thermal contribution to the fermionic self-energy is dominant: $\Sigma(i\omega_n)\approx g^2 T \mathcal{D}(0)\mathcal{G}(i\omega_n)=\lambda\Lambda TL(0)/(i\omega_n-\Sigma(i\omega_n))$, (where $L(i\Omega_m)$ is defined in Eq.~\eqref{eq:logarithm}). Solving for the fermionic self-energy, we arrive at the impurity-like NFL solution~\cite{EsterlisSchmalian2019} to Eq.~\eqref{eq:Sigma_saddlept}:
\begin{equation}
\Sigma(i\omega_n)=-i\text{sgn}(\omega_n)\frac{1}{2}\left[\sqrt{\omega_n^2+\Omega_0^2}-|\omega_n|\right]\sim -i\text{sgn}(\omega_n)\frac{\Omega_0}{2}.
\label{eq:impurity_self-energy}
\end{equation}
The fermionic energy scale $\Omega_{0}$ is defined by $\Omega_0\equiv 2\sqrt{\lambda\Lambda T L(0)}$, where we have adjusted the units to match those of the leading-order expansion in $\Lambda$ of the higher-dimensional theory. In this regime, the bosons are underdamped and act as classical impurities for electrons, which are described by the effective Green's function
\begin{equation}
\mathcal{G}(i\omega_n) = -\frac{2i\nu\Lambda \text{sgn}(\omega_n)}{\sqrt{\omega_n^2+\Omega_0^2}+|\omega_n|}.
\label{eq:impurity-G}
\end{equation}

The bosonic self-energy can be evaluated using the impurity-like solution for $\Sigma(i\omega_n)$. In contrast to the infinite-bandwidth limit, an additional constraint  is not required to tune the system to the quantum critical point. Rather, the solution is found to ``self-tune'' to criticality through the electron-boson interaction for an arbitrary bare bosonic mass, $M_0$. This result is known within the context of the (0+1)-dimensional Y-SYK model~\cite{EsterlisSchmalian2019, Wang2020, ClassenChubukov2021, Pan2021}. Here, we recover this result for the (2+1)-dimensional model in the impurity regime. By evaluating the renormalized boson mass ($M^2\equiv M^{2}_0-\Pi(0)$) in terms of the static bosonic self-energy, we find
\begin{align}\nonumber
M^2(T)=M_0^2-\Pi(0,T) &=M_0^2 +2(g\nu\Lambda)^2 T\sum_{\omega_n}\left(\frac{1}{i\omega_n-\Sigma(i\omega_n)}\right)^2\\ 
&=M_0^2-8(g\nu\Lambda)^2 T \sum_{\omega_n}\left(\frac{1}{\sqrt{\omega_n^2+\Omega_0^2}+|\omega_n|}\right)^2.
\label{eq:BosonMassFiniteT1}
\end{align}
For $T\ll \Omega_0$, the summation is evaluated as
\begin{equation}
M^2=M_0^2-8(g\nu\Lambda)^2\int_{0}^{\infty}\frac{1}{\left(\sqrt{\omega^2+\Omega_0^2}+\omega\right)^2}\frac{d\omega}{\pi} = M_0^2-\frac{16(g\nu\Lambda)^2}{3\pi\Omega_0} = M_{0}^{2}\left(1 -\frac{8}{3\pi \kappa L(0)} \frac{\Omega_{0}}{T}\right).
\label{eq:M0}
\end{equation}
We define $\kappa\equiv2M_{0}^2\varepsilon_{F}/(\omega_{D}^2\Lambda)$. Substituting our expression for $\Omega_0$, we obtain a self-consistent equation for the renormalized boson mass
\begin{equation}
M^2 = M_0^2 - \frac{8\omega_D^2}{3\pi}\frac{\Lambda}{\varepsilon_F}\sqrt{\frac{\lambda\Lambda/T}{\log(1+\omega_D^2/M^2)}}.
\end{equation}
For $M/\omega_{D}\ll1$, the solution for the renormalized boson mass is
\begin{equation}
\frac{M}{\omega_{D}}\sim \exp\left(-\frac{128\lambda}{9\pi^2 \kappa^2}\frac{\Lambda}{\omega_{D}}\frac{\omega_{D}}{T}\right).
\label{eq:RenormalizeBosonMass_finiteBW}
\end{equation}
This solution is indeed consistent with the property that $M/\omega_{D}\ll1$ for $T/\omega_{D}\ll1$. 
For comparison, the low-temperature renormalized boson mass for the (0+1)-dimensional model in the impurity regime is determined by $M^2=M_0^2-8gM/(3\pi\sqrt{T})$ (with units adjusted from Ref.~\cite{EsterlisSchmalian2019} to match our own). The renormalized boson mass for the 2d model takes on a different form compared to that of Ref.~\cite{EsterlisSchmalian2019} due to the boson being linearly dispersive in 2d. We compare the low-temperature solutions for the (2+1)d and (0+1)d models in Fig.~(2) of the main text. Finite-temperature corrections to the summation in Eq.~\eqref{eq:BosonMassFiniteT1} are evaluated with the Abel-Plana formula~\cite{OlverBook}. In terms of the dimensionless temperature $T/\Omega_0$, the self-consistent equation for the renormalized boson mass is 
\begin{equation}
\left(\frac{M}{M_{0}}\right)^2 = 1 -\frac{8}{3\pi \kappa L(0)} \frac{\Omega_{0}}{T} \left[1-\frac{\pi^2}{2}\left(\frac{T}{\Omega_{0}}\right)^2+\dots\right].
\label{eq:Mass-FBW-Tcorr}
\end{equation}

\begin{figure}[ht]
\centering\includegraphics[width=.5\columnwidth]{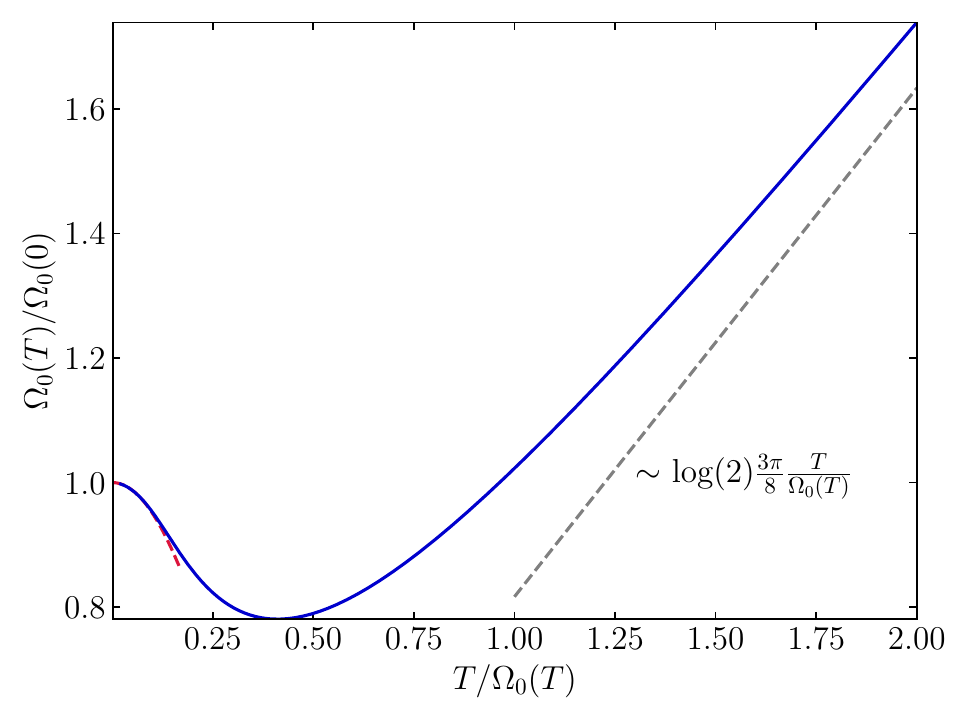}
\caption{Electronic energy scale for the impurity regime normalized by its zero-temperature value. The blue curve denotes the numerically evaluated finite-temperature value of the impurity regime energy-scale. The dashed red and gray curves are the low and high-temperature asymptotic values respectively. Here, $\omega_D=M_0=\varepsilon_F=\Lambda/2$.}
\label{fig:FBW-energy-scale-T-dep}
\end{figure}

The dynamical portion of the bosonic self-energy in the impurity regime can be determined by using the approximate low-energy expression for the electronic Green's function:
\begin{equation}
\Pi(i\Omega_{m})-\Pi(0) \approx 8\left(\frac{\nu\Lambda g}{\Omega_0}\right)^2 T \sum_{\omega_n}\left(\text{sgn}(\omega_n)\text{sgn}(\omega_n+\Omega_m)-1\right) = -\frac{8}{\pi}\left(\frac{\nu\Lambda g}{\Omega_0}\right)^2|\Omega_m|. 
\end{equation}
Substituting this expression for the renormalized boson mass into the bosonic self-energy, we find $\Pi(i\Omega_{m})-\Pi(0)\approx-9\pi \kappa^2 \omega_D^2|\Omega_m|/(128\lambda \varepsilon_{F})$, which can be neglected for large enough coupling. At low temperatures and in the infrared (IR), the impurity scattering rate is given by
\begin{equation}
\frac{1}{2\tau} \sim\frac{1}{2}\Omega_0 \equiv  \frac{8\omega_D^2\Lambda^2}{3\pi M_0^2\varepsilon_{F}}\lambda=\frac{16}{3\pi}\frac{\lambda}{\kappa}\Lambda.
\label{eq:Omega0}
\end{equation}
The requirement $\Omega_0\gg\Lambda$ is equivalent to $\lambda\gg(3\pi/32)\kappa$. For equal electronic and bosonic energy scales: $\omega_D=M_0=\varepsilon_{F}=\Lambda/2$, the parameter $\kappa=1$, and the separation of scales required for the impurity-regime is $\lambda\gtrsim1$~\footnote{The error in approximating the arctangent by the first term of its series can be estimated for small values of coupling as follows; let $\kappa=\lambda =1$, $\text{arctan}(3\pi/32)\simeq3\pi/32$, with an error of approximately 3\%. In the limit $\lambda/\kappa\rightarrow\infty$, $\text{arctan}(x)\rightarrow x$.}. The energy scale $\Omega_0$ is shown in Fig.~\ref{fig:FBW-energy-scale-T-dep} for arbitrary temperatures.

We now determine the low-temperature threshold for the impurity regime by using an IR analysis. For $|\omega_n|\ll\Omega_0$, $\mathcal{G}(i\omega_n)\sim-2i\nu\Lambda\text{sgn}(\omega_n)/\Omega_0$ and thus the fermionic self-energy becomes
\begin{equation}
\Sigma(i\omega_n)=-i\text{sgn}(\omega_n)\sqrt{\frac{\lambda \Lambda T}{L(0)}}\left[L(0)+2\sum_{m=1}^{|n|}L(i\Omega_{m})\right] = -i\text{sgn}(\omega_n)\left[\frac{\Omega_0}{2}+2\sqrt{\frac{\lambda \Lambda T}{L(0)}}\sum_{m=1}^{|n|}L(i\Omega_{m})\right].
\end{equation}
The form of the electronic self-energy in the impurity regime assumes that $L(0)\gg2\sum_{m=1}^{|n|} L(i\Omega_{m})$; that is, the thermal (resonant) contribution to the fermionic self-energy is dominant to the quantum (non-resonant) contribution. For sufficiently strong coupling, the dynamical part of the bosonic self-energy can be neglected and $M\ll T$; thus $L(i\Omega_{m})\approx \log(1+\omega_D^2/(M^2+\Omega_m^2))\approx \log(1+\omega_D^2/\Omega_m^2)$ for $\Omega_m\neq0$. Let $\widetilde{\omega}_{D}\equiv\omega_{D}/(2\pi T)$. The non-resonant contribution to the fermionic self energy is then given by
\begin{align}
\sum_{m=1}^{|n|}L(i\Omega_{m}) &= \sum_{m=1}^{|n|}\log\left(1+\omega_D^2/\Omega_m^2\right) =\int_{0}^{1}\left[\sum_{m=1}^{|n|}\frac{\widetilde{\omega}_{D}^2}{m^2+x\widetilde{\omega}_{D}^2}\right]dx \nonumber\\
&=\int_{0}^{1} \frac{i\widetilde{\omega}_{D}}{2\sqrt {x}}
\left[-\psi \left( n+1-i\sqrt {x}\widetilde{\omega}_{D} \right) + \psi \left( n+1+i\sqrt {x}\widetilde{\omega}_{D} \right) 
+ \psi \left( 1-i\sqrt {x}\widetilde{\omega}_{D} \right) - \psi \left( 1+i\sqrt {x}\widetilde{\omega}_{D} \right)\right] dx\nonumber\\
&=2\log\left(\left|\frac{\Gamma\left(|n|+1+i\widetilde{\omega}_{D}\right)}{|n|!\Gamma\left(1+i\widetilde{\omega}_{D}\right)}\right|\right).
\end{align}
The bosonic UV cutoff $\omega_D$ plays a similar role to the electronic bandwidth. Consequently, for $\omega_D/T\ll1$, the form of the bosonic propagator in the (0+1)d model~\cite{EsterlisSchmalian2019} (for $m\neq0$) is recovered:
\begin{equation}
\sum_{m=1}^{|n|}L_m \approx \sum_{m=1}^{|n|}\frac{\omega_D^2}{\Omega_m^2}=\frac{\omega_D^2}{(2\pi T)^2}H_{|n|}^{(2)},
\end{equation}
where $H_n^{(2)}$ is the $n$-th harmonic number of the second order. The thermal (resonant) contribution to the sum is $L(0)\sim (8/(3\pi\kappa))(\Omega_0/T)\simeq (256/(9\pi^2))(\lambda/\kappa^2)(\Lambda/T)$, using the low-temperature ($T\ll \lambda\Lambda/\kappa^2$) value of the renormalized boson mass in Eq.~\eqref{eq:M0}. The boson mass and self-energy only weakly contribute to the sum and further depreciate the quantum contribution to the electronic self-energy. The separation of scales required for the impurity regime are satisfied for $T>T^*$ and $\lambda\gg\kappa$. By demanding $L(0)\gg 2\sum_{m=1}^{|n|} L_m$, we obtain the lower temperature thresholds for the impurity regime: 
\begin{align}\label{eq:thresh1}
\lambda\Lambda/\kappa^2 &\gg T^*\log(\omega_D/T^*);\quad T\ll \omega_D,\\ \label{eq:thresh2}
T^{*}&\gg \kappa^2\omega_D^2/(\lambda\Lambda);\;\;\;\;\;\quad T\gg \omega_D,
\end{align}
Assuming non-adiabatic units and strong coupling, the temperature threshold for $\Sigma^{\text{th}}\gg\Sigma^{\text{qm}}$ is automatically satisfied.

Surprisingly, we find the thermal contribution to the fermionic self-energy to be non-vanishing and dominant in the strong-coupling regime down to arbitrarily low temperatures. This is evident from the scaling of $\Sigma^{\text{th}}\propto \text{const.}$ and $\Sigma^{\text{qm}}\propto T\log(\omega_D/T)$ at the lowest temperatures. In contrast, in the (0+1)d Y-SYK model, the impurity regime terminates at the low-temperature cutoff $\sim (M_0^2/g)^2$ (analogous to Eq.~\eqref{eq:thresh2}), crossing over to an SYK-like non-Fermi-liquid regime with a distinct universal exponent~\cite{EsterlisSchmalian2019}. Equation~\eqref{eq:thresh1} indicates that an analogous low-temperature fixed point within the strong-coupling regime of the (2+1)d Y-SYK model does not exist at nonzero temperatures.

This difference from the lower-dimensional theory can be attributed to the bosons in our analysis being linearly dispersive in 2d. Even in the strong-coupling limit, the bosonic self-energy remains an insignificant energy scale with respect to the bosonic UV cutoff, $\omega_D$, preventing the logarithm from being expanded as in Eq.~\eqref{eq:D-eff-expanded}. By retaining the logarithm, the low-temperature scaling of the quantum contribution of the fermionic self-energy is softened, allowing for the resonant contribution to be dominant at arbitrarily low temperatures.

\begin{figure}[ht]
\centering\includegraphics[width=.5\columnwidth]{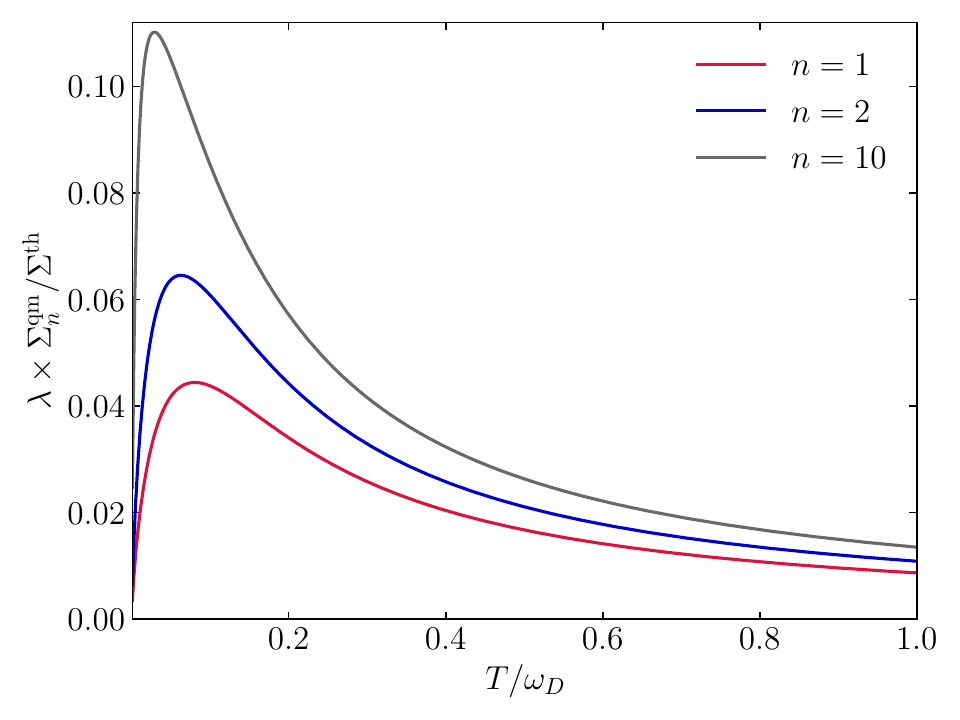}
\caption{Ratio of quantum and thermal contributions to the electronic self-energy within the impurity regime multiplied by dimensionless coupling $\lambda$. Landau damping in the bosonic propagator is neglected and units are set to the non-adiabatic values $\omega_D=M_0=\varepsilon_{F}=\Lambda/2$. The provided IR analysis is applicable for $|n|\ll \lambda\omega_D/T$ (in non-adiabatic units). The relative weight of the quantum to thermal contributions is a maximum near $T_c$.}
\label{fig:Ratio}
\end{figure}

\subsubsection*{Ultra-high temperature}

For $T \gg \Lambda/(2\pi)$, the separation of scales $|i\omega_n-\Sigma(i\omega_n)| \gg \Lambda/2$ is guaranteed. The asymptotic behavior of the momentum-integrated electronic Green's function is $\mathcal{G}(i\omega_n)\sim-i\nu\Lambda/\omega_n$. The self-energies for the electron and boson are suppressed at high temperatures, $\Sigma(i\omega_n)\sim -i \Omega_0^2/(4\omega_n)$, thus
\begin{align}
\Pi(i\Omega_m) &\sim 2(g\nu\Lambda)^2T\sum_{\omega_n}\frac{1}{\omega_n(\omega_n+\Omega_m)} = \frac{2T}{\Omega_{m}}(g\nu\Lambda)^2\sum_{n=0}^{\infty}\left(\frac{1}{\omega_{n}-\Omega_{m}}-\frac{1}{\omega_{n}+\Omega_{m}}\right)  \nonumber\\
&=\frac{(g\nu\Lambda)^2}{\pi \Omega_{m}}\left[\psi\left(\frac{1}{2}+\frac{\Omega_{m}}{2\pi T}\right) - \psi\left(\frac{1}{2}-\frac{\Omega_{m}}{2\pi T}\right)\right] \nonumber\\ &=\frac{(g\nu\Lambda)^2}{\Omega_{m}}\tan\left(\frac{\Omega_{m}}{2T}\right)
=\delta_{m,0}\frac{(g\nu \Lambda)^2}{2T}.\label{eq:Pi-ultra-high-T}
\end{align}
Setting the bosonic Matsubara frequency to zero, the renormalized boson mass is $M^2\sim M_0^2-(g\nu\Lambda)^2/(2T)$ for temperatures $T\gg\Omega_0$, approaching its bare value for sufficiently high temperatures. Thus, for $T\gg\Omega_0$, $\Omega_0(T)\sim 2\sqrt{\lambda\Lambda T \log(1+\omega_D^2/M_0^2)}$. In the crossover from low to ultra-high temperatures, the electronic self-energy may be considered, which provides dynamical corrections to Eq.~\eqref{eq:Pi-ultra-high-T}.

\section{Entropy}

\subsection{Entropy in the infinite-bandwidth limit}

In the normal phase, the Yukawa-SYK action (including the constraint) is
\begin{align} 
S &=-2 N \int_k \log\left(-i\omega_n + \xi_{\bm k} + \Sigma(k)\right) + \frac{N}{2} \int_q \log\left(\Omega_m^2 +c^2\bm{q}^2 +M_0^2- \Pi(q)\right) - N \int \frac{i\lambda(\tau,\bm{r})}{2\gamma}d\tau d^2r \nonumber \\
&\quad -2 N \left[\int_k  \Sigma(k) G(k) -\frac{1}{4}\int_q  \Pi(q) D(q) - \frac{g^2}{2} \int_{k,k',\bm{q}} G(k) D(i\omega_n-i\omega_{n^\prime},\bm{q}) G(k') \right].
\label{eq:S}
\end{align}
This expression has the same general form as in Luttinger and Ward's~\cite{Luttinger1960} and Eliashberg's papers~\cite{Eliashberg1963}. At the saddle-point level, the second and third terms in square brackets cancel each other due to the variational condition $\delta S/\delta D=0$. In particular, both terms describe the exact same skeleton diagram in the free energy~\cite{Zhang2023}. Thus, only the first term in the square brackets contributes to the saddle-point action. 

Using the definitions $\mathcal{G}(i\omega_{n})\equiv \int_{\bm{k}} G(k)$, $\mathcal{D}(i\Omega_{m})\equiv \int_{\bm{q}} D(q)$, and $\lambda=\nu g^2 /\left(4 \pi c^2\right)$ and substituting the saddle-point self-energies from Eqs.~\eqref{eq:Sigma_saddlept} and~\eqref{eq:Pi-saddlept} into the action, along with the mass constraint $i\lambda(\tau,\bm{r})=M_0^2$, we obtain the normal-state grand canonical potential:
\begin{equation}
\mathcal{F} = N\left[-2T\int_{\bm{k}}\sum_{\omega_n} \log (-G^{-1}(k)) -2 T \sum_{\omega_n}\mathcal{G}(i\omega_n)\Sigma(i\omega_n)+ \frac{1}{2} T \int_{\bm{q}}\sum_{\Omega_m} \log (D^{-1}(q)) -\frac{M^2(T)}{2\gamma}\right]. \label{eq:FreeEnergy}
\end{equation}

\subsubsection*{Fermionic entropy}

Next, we add and subtract the free-fermion free energy, which then allows the Matsubara frequency summation and momentum integration to be interchanged~\cite{AGD}. The fermionic contribution is then given by~\cite{Luttinger1960}:
\begin{align}
\mathcal{F}_{F} &= N\left[-2\int_{\bm{k}}T\sum_{\omega_n} \log (-G^{-1}(k))  -2 T \sum_{\omega_n}\mathcal{G}(i\omega_n)\Sigma(i\omega_n)\right] \nonumber \\ 
&= -2NT\left[\int_{\bm{k}}T\sum_{\omega_n} \log \left(\frac{G^{-1}(k)}{G^{-1}_{0}(k)}\right) + \int_{\bm{k}}T\sum_{\omega_n} \log (-G^{-1}_0(k)) + \sum_{\omega_n}\mathcal{G}(i\omega_n)\Sigma(i\omega_n)\right].
\end{align}
The first integral can be evaluated as follows:
\begin{align}
-\int_{\bm{k}} \log \left(\frac{G^{-1}(k)}{G^{-1}_{0}(k)}\right)&=-\int_{\bm{k}}\log(1-G_{0}(k)\Sigma(i\omega_n)) \nonumber\\
&=\int_{\bm{k}}\sum_{m=1}^{\infty}\frac{1}{m}(G_{0}(k)\Sigma(i\omega_n))^m \nonumber \\ 
&=\sum_{m=1}^{\infty}\frac{1}{m}(\Sigma(i\omega_n))^m\int_{\bm{k}}(G_{0}(k))^m =  \Sigma(i\omega_n)\mathcal{G}(i\omega_n).
\end{align}
The last line is valid only in the infinite bandwidth limit, since $\int_{\bm{k}}(G_{0}(k))^m=\nu \int_{-\infty}^{\infty} (i\omega_n-\xi)^{-m}d\xi=0$ for $m>1$. Thus, for the fermionic free energy, the first term cancels the $\mathcal{G}\Sigma$ term, as also found in Refs.~\cite{Esterlis2021, Zhang2023}. This result is true for a local fermionic self energy~\cite{Zhang2023}. We are left with only the free contribution to the fermionic free energy 
\begin{equation}
\mathcal{F}_{F}=-2NT\int_{\bm{k}}\log\left(1+e^{- \xi_{\bm{k}}/T}\right).
\end{equation}
The fermionic entropy-density is given by $s_{F} = -\frac{1}{N}(\partial_T \mathcal{F}_F)_{V,\mu}$, thus
\begin{align} 
s_{F}&=2\int_{\bm{k}}\left[ \frac{\xi_{\bm k}}{T}n_F(\xi_{\bm k})-\log\left(1-n_F(\xi_{\bm k})\right)\right] \nonumber\\
    &=-2\int_{\bm{k}}\left[ n_F(\xi_{\bm k})\log\left(n_F(\xi_{\bm k})\right)+(1-n_F(\xi_{\bm k}))\log\left(1-n_F(\xi_{\bm k})\right)\right],
\label{Free-fermion-entropy1}
\end{align}
where $\xi/T=\log\left(\frac{1-n_F(\xi)}{n_F(\xi)}\right)$ was used. Expressed as a $\xi$-integral, the fermionic entropy density is
\begin{align}
s_{F}&=-2 \nu\int_{-\infty}^{\infty} \left[n_F(\xi)\log\left(n_F(\xi)\right)+(1-n_F(\xi))\log\left(1-n_F(\xi)\right)\right] d\xi \nonumber\\
&=-2T \nu\int_{-\infty}^{\infty} \left[f(y)\log\left(f(y)\right)+(1-f(y))\log\left(1-f(y)\right)\right] dy; \quad y = \xi/T,f(y)=\frac{1}{1+e^y} \nonumber\\
&= \frac{2\pi^2 }{3}\nu T.
\label{eq:FermionEntropy}
\end{align}
After accounting for spin, this result agrees with the Sommerfeld result.

\subsubsection*{Bosonic entropy}

The bosonic contribution to the free energy is
\begin{equation}
\mathcal{F}_B = \frac{N}{2}\left[T\int_{\bm{q}}\sum_{\Omega_m} \log (D^{-1}(q)) - \frac{1}{\gamma}M^2(T)\right]. 
\label{eq:F_boson}
\end{equation}
Entropy is determined from $s_{B} = -\frac{1}{N}(\partial_T \mathcal{F}_B)_{V,\mu}$. Performing the Matsubara frequency summation using a contour integral, and using the constraint equation in Eq.~\eqref{eq:Constraint_saddlept}, the bosonic contribution to the entropy is
\begin{align}
s_{B} &= -\frac{1}{2}\left[\partial_T \int_{\bm{q}}\int_{-\infty}^{\infty}n_B(\Omega)\text{Im}\log\left(D_R^{-1}(\Omega,\bm{q})\right)\frac{d\Omega}{\pi}-\frac{\partial_T M^2(T)}{\gamma}\right] \nonumber \\
 &= -\frac{1}{2}\left\{ \int_{\bm{q}}\int_{-\infty}^{\infty}\left[\left(\partial_T n_B(\Omega)\right)\text{Im}\log\left(D_R^{-1}(\Omega,\bm{q})\right)+(\partial_T M^2(T))n_B(\Omega)\text{Im}D_R(\Omega,\bm{q})\right]\frac{d\Omega}{\pi} - \partial_T M^2(T)\frac{1}{\gamma}\right\} \nonumber \\
  &= -\frac{1}{2\pi}\int_{\bm{q}}\int_{-\infty}^{\infty}\left(\partial_T n_B(\Omega)\right)\text{Im}\log\left(D_R^{-1}(\Omega,\bm{q})\right) d\Omega. 
 \label{eq:BosonEntropy1}
\end{align}
Note that the temperature derivative now acts only on the distribution function; an analogous result is found in Sec. 19.5 of Ref.~\cite{AGD}. Substituting $c^2q^2=x\omega_{D}^2$, Eq.~\eqref{eq:BosonEntropy1} can then be expressed as 
\begin{equation}
s_{B} = -\frac{\nu\varepsilon_{F}}{2\pi}\int_0^{1}\left[\int_{-\infty}^{\infty}  \left(\partial_T n_B(\Omega)\right) \text{Im}\log\left(-\Omega^2+x\omega_{D}^2+M^2-2\pi i\lambda_E \omega_D\Omega\right)d\Omega\right]dx.
\end{equation}
Here we used $1/(4\pi c^2)=\nu\varepsilon_{F}/\omega_D^2$. Next, we carefully evaluate the imaginary part of the logarithm, accounting for which quadrant of the complex plane the logarithm lies in, and find that the entropy becomes
\begin{equation}
s_{B} = \frac{\nu\varepsilon_{F}}{2\pi}\int_0^{1}\left[\int_{-\infty}^{\infty} \left(\partial_T n_B(\Omega)\right)\arctan\left(\frac{2\pi\lambda_{E}\omega_{D}\Omega}{-\Omega^2+x\omega_{D}^2+M^{2}}\right) 
d\Omega \right]dx + \nu\varepsilon_{F}\int_0^{1}\left[\int_{\sqrt{x\omega_{D}^2+M^2}}^{\infty} \left(\partial_T n_B(\Omega)\right)
d\Omega \right]dx.
\label{eq:BosonEntropy2}
\end{equation}
If we set $\lambda=0$, then the first term vanishes. Thus, the second term can be recognized as the entropy density for a massive free boson in (2+1)d. Note that the mass is held fixed, since the temperature dependence of the mass was shown above to cancel with the constraint term. Let $\omega_{B}=\sqrt{\omega_D^{2}+M^2}$. Evaluating the integral over $\Omega$, we obtain 
\begin{align}
s_{B,0} &= -\nu\varepsilon_{F}\partial_T\left[\int_0^{1}\log\left(1-\exp\left(-\sqrt{x\omega_{D}^2+M^2}/T\right)\right)dx\right] \nonumber\\
&= -\frac{2\nu\varepsilon_{F}}{\omega_D^2}\partial_T\left[T^3\int_{M/T}^{\omega_B/T}y\log\left(1-\exp(-y)\right)dy\right].
\end{align}
To evaluate this integral, we use the polylogarithm function, which is defined by $\text{Li}_{s}(z)=\sum_{k=1}^{\infty}z^k/{k^s}$~\cite{LewinBook}. 
\begin{align}
\int_{M/T}^{\omega_B/T}y\log(1-\exp(-y))dy & =-\sum_{k=1}^{\infty}\frac{1}{k}\int_{M/T}^{\omega_B/T}y\exp(-yk)dy \nonumber\\ 
&= \sum_{k=1}^{\infty}\left[\left.\frac{(yk+1)\exp(-yk)}{k^{3}}\right|_{M/T}
^{\omega_{B}/T}\right] \nonumber\\ 
&= \textrm{Li}_{3}(e^{-\omega_{B}/T})-\textrm{Li}_{3}(e^{-M/T})+\frac{\omega_{B}}{T}\textrm{Li}_{2}(e^{-\omega_{B}/T})-\frac{M}{T}\textrm{Li}_{2}(e^{-M/T}).
\end{align}
Using the above result, the entropy reduces to
\begin{align}
s_{B,0}	&= 2\nu\varepsilon_{F}\biggl\{\left(\frac{\omega_{B}}{\omega_{D}}\right)^{2}\log\left(1-e^{-\omega_{B}/T}\right)-\left(\frac{M}{\omega_{D}}\right)^{2}\log\left(1-e^{-M/T}\right) \nonumber\\
&\quad +3\left(\frac{T}{\omega_{D}}\right)^{2}\left[\frac{M}{T}\textrm{Li}_{2}(e^{-M/T})-\frac{\omega_{B}}{T}\textrm{Li}_{2}(e^{-\omega_{B}/T})+\textrm{Li}_{3}(e^{-M/T})-\textrm{Li}_{3}(e^{-\omega_{B}/T})\right]\biggr\}.
\label{eq:boson-free-entropy}
\end{align}
The free-boson contribution to the entropy density has the following low- and high-temperature expansions:
\begin{equation} 
s_{B,0} = \frac{\nu\varepsilon_{F}\omega_B^2}{\omega_D^2}
\begin{cases}
6\zeta(3)\left(\frac{T}{\omega_B}\right)^2;\quad T\ll \omega_B\\
\log\left(\frac{T}{\omega_B}\right);\hspace{8mm} T \gg \omega_B
\end{cases} 
\label{eq:impurity-entropy}
\end{equation}
In the above results, we have used the fact that $M(0)=0$. 

\begin{figure}[t]
\centering\includegraphics[width=.5\columnwidth]{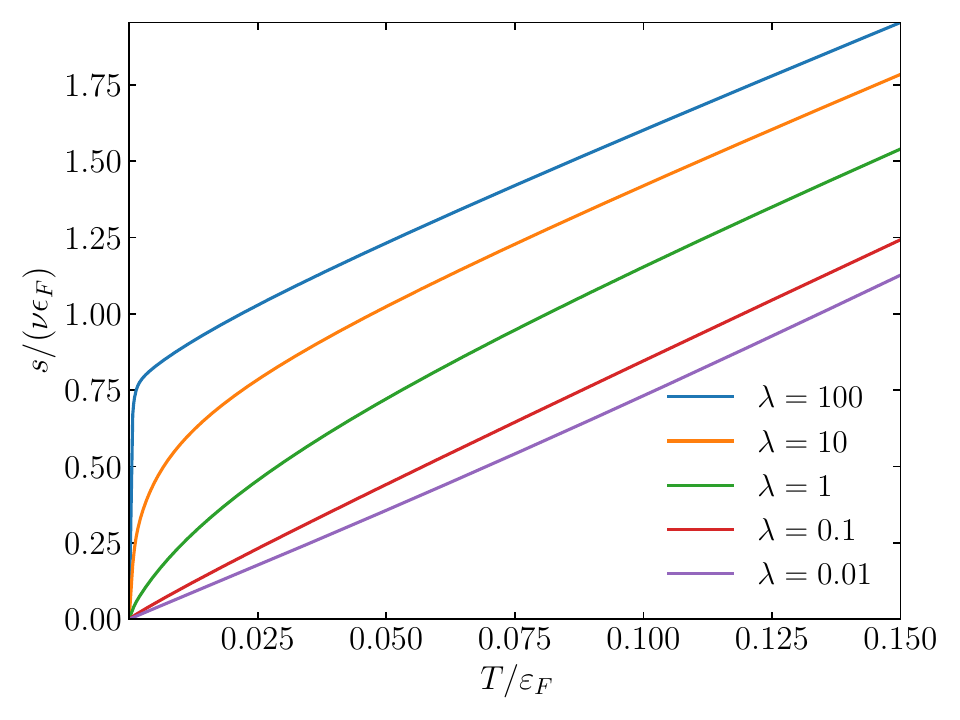}
\caption{Entropy density in the infinite bandwidth limit in units of $\nu\varepsilon_{F}$ as a function of temperature for various values of the dimensionless coupling strength $\lambda$. The Fermi energy and bosonic UV cutoff are set to the non-adiabatic values $\varepsilon_F=\omega_D$.}
\label{fig:Entropy}
\end{figure}

The contribution to the entropy arising from the electron-boson interaction is given by the first term in Eq.~\eqref{eq:BosonEntropy2}. For low temperatures and weak coupling, we can approximate this contribution to the entropy as follows:
\begin{align}
s_{\text{Int}}&=
\frac{\nu\varepsilon_{F}}{2\pi}\int_{0}^{1}\left[\int_{-\infty}^{\infty} y\ \text{cosech}^{2}(y) \arctan\left(\frac{4\pi\lambda_{E}\omega_{D}Ty}{-(2Ty)^2+x\omega_{D}^2+M^{2}}\right) 
dy \right]dx \nonumber \\
&\approx \frac{\nu\varepsilon_{F}}{2\pi}4\pi\lambda_{E}T\omega_{D}\int_{0}^{1}\frac{1}{x\omega_{D}^2 + M^2}dx \int_{-\infty}^{\infty}y^2 \text{cosech}^2(y) dy \nonumber\\ 
&=\frac{2\pi^2}{3} \frac{T}{\omega_{D}} \lambda_{E}\nu\varepsilon_{F} \log\left(1+\frac{\omega_D^2}{M^2}\right).
\end{align}
The integral over $y$ can be evaluated using the digamma function and gives $\pi^2/3$, and the integral over $x$ is elementary. Combining the electronic contribution with the low-temperature bosonic contribution, we find that the entropy density has the expected $\sim T\log T$  form as in a marginal Fermi liquid~\cite{Crisan1996}:
\begin{equation}
s_{F,0}+s_{\text{Int}}\approx \frac{2\pi^2}{3} \lambda \nu T\log\left(e^{1/\lambda}\frac{\omega_D^2}{M^2}\right).
\end{equation}
The total entropy (fermionic plus bosonic) is determined by numerically evaluating the sum of Eqs.~\eqref{eq:FermionEntropy} and \eqref{eq:BosonEntropy1}; the result is shown in Fig.~\ref{fig:Entropy}. At high temperatures, the fermionic contribution dominates and $s\propto T$. However, at low temperatures and intermediate coupling $s\sim T\log T$, which is indicative of marginal-Fermi-liquid behavior.

\subsection{Entropy in the finite-bandwidth limit}

In this section, we evaluate the fermionic contribution to the free energy using the impurity-like solution in Eq.~\eqref{eq:impurity_self-energy}, which arises in the finite-bandwidth limit. The first term in the free energy can be expressed as~\cite{MahanBook}:
\begin{align}
2NT\int_{\bm{k}}\sum_{\omega_n} \log (-G^{-1}(k)) &= 2NT\int_{\bm{k}}\biggl\{\lim_{\delta\rightarrow0}\sum_{\omega_n} \exp(i\omega_{n}\delta) \log (-i\omega_n + \xi_{\bm{k}}+ \Sigma(i\omega_n))\biggr\} \nonumber\\
&=-\frac{2}{\pi}N\int_{\bm{k}}\int_{-\infty}^{\infty} n_F(\omega)\text{Im}\left[\log (-\omega + \xi_{\bm{k}}+ \Sigma_{R}(\omega)) \right] d\omega. 
\end{align}
The contribution to the entropy density is then given by
 \begin{align}\nonumber
s_{G} &= \frac{2}{\pi}\int_{\bm{k}}\int_{-\infty}^{\infty}\left(-\frac{\partial n_F(\omega)}{\partial T}\right)\text{Im}\left[\log (-G_R^{-1}(\omega,\bm{k}))\right] d\omega\\ \nonumber
 &= -\frac{2}{\pi T}\int_{\bm{k}}\int_{-\infty}^{\infty}\left(-\frac{\partial n_F(\omega)}{\partial \omega}\right)\omega\;\text{Im}\left[\log(-G_R^{-1}(\omega,\bm{k}))\right] d\omega\\
 &\sim-\frac{2\pi}{3}\nu T \int_{-\Lambda/2}^{\Lambda/2} \text{Im}\left.\left[G_R(\omega,\xi)\partial_{\omega}G^{-1}_R(\omega,\xi)\right]\right|_{\omega=0} d\xi. 
\end{align}
In the last step, we used the Sommerfeld expansion~\cite{AGD}: $\int_{-\infty}^{\infty} f(\varepsilon) \frac{\partial n_F(\varepsilon)}{\partial \varepsilon} d\varepsilon=-f(0)-\frac{\pi^2}{6} T^2\left.\frac{\partial^2 f}{\partial \varepsilon^2}\right|_{\varepsilon=0}+\cdots$ to evaluate the entropy for $T\ll \Omega_0$, where variations $\partial_T \Omega_0(T)$ can be ignored. Analytically continuing Eq.~\eqref{eq:impurity_self-energy}, $\Sigma_R(\omega)=-\frac{i}{2}\left[\sqrt{\Omega_0^2-\omega^2}+i\omega\right]$ for $|\omega|<\Omega_0$ within the impurity regime. Thus,
\begin{equation}
s_{G} = \frac{\pi}{3}\nu T\int_{-\Lambda/2}^{\Lambda/2}\text{Im}G_{R}(0,\xi)d\xi = \frac{2\pi}{3}\nu T\int_{-\Lambda/2}^{\Lambda/2} \frac{\Omega_0}{\Omega_0^2+4\xi^2} d\xi \nonumber = 
\frac{2\pi}{3}\nu  T \text{arctan}\left(\Lambda/\Omega_0\right)\sim \frac{2\pi\nu \Lambda T}{3\Omega_0}.
\end{equation}
In the last step, we invoked the separation of scales $\Omega_0\gg \Lambda$.

Next, we evaluate the contribution to the free energy from the $\mathcal{G}$-$\Sigma$ interaction using Eqs.~\eqref{eq:impurity_self-energy}-\eqref{eq:impurity-G}: 
\begin{equation}
\mathcal{F}_{\mathcal{G}\Sigma} = -2N T \sum_{\omega_n}\mathcal{G}(i\omega_n)\Sigma(i\omega_n)
=4N\nu\Lambda T\sum_{n=0}^{\infty}\frac{\sqrt{\omega_n^2+\Omega_0^2}-\omega_n}{\sqrt{\omega_n^2+\Omega_0^2}+\omega_n}.
\end{equation}
To obtain the large-$\Omega_{0}/T$ behavior of this expression, we apply the Abel-Plana formula~\cite{OlverBook}:
\begin{equation}
\sum_{n=0}^{\infty}f(n)=\frac{1}{2}f(0)+\int_{0}^{\infty}f(x)dx+i\int_{0}^{\infty}\frac{f(it)-f(-it)}{e^{2\pi t}-1}dt.
\end{equation}
Applying this formula to the Matsubara frequency summation, we have 
\begin{align}
2\sum_{n=0}^{\infty}\frac{\sqrt{\omega_{n}^{2}+\Omega_{0}^{2}}-\omega_{n}}{\sqrt{\omega_{n}^{2}+\Omega_{0}^{2}}+\omega_{n}} 
&= \frac{\sqrt{(\pi T)^{2}+\Omega_{0}^{2}}-\pi T}{\sqrt{(\pi T)^{2}+\Omega_{0}^{2}}+\pi T}+2\int_{0}^{\infty}\frac{\sqrt{(2x+1)^{2}(\pi T)^{2}+\Omega_{0}^{2}}-(2x+1)\pi T}{\sqrt{(2x+1)^{2}(\pi T)^{2}+\Omega_{0}^{2}}+(2x+1)\pi T}dx \nonumber\\
& \quad +2i\int_{0}^{\infty}\frac{f(it)-f(-it)}{e^{2\pi t}-1}dt.
\end{align}
For convenience, we do not explicitly write out the third term. Performing the integrals over $x$ and $t$, and taking the limit $\Omega_{0}/T\gg1$, we find: 
\begin{equation}
2T\sum_{n=0}^{\infty}\frac{\sqrt{\omega_{n}^{2}+\Omega_{0}^{2}}-\omega_{n}}{\sqrt{\omega_{n}^{2}+\Omega_{0}^{2}}+\omega_{n}}=\frac{2\Omega_{0}}{3\pi}-\frac{\pi}{3\Omega_{0}}T^{2}+\dots.
\end{equation}
Therefore, the entropy density is given by $s_{\mathcal{G}\Sigma}=4\pi\nu\Lambda T/(3\Omega_0)$. Combining both fermionic contributions, the total fermionic entropy is 
\begin{equation}
s_{F}=s_{G}+s_{\mathcal{G}\Sigma}=2\pi\nu\Lambda \frac{T}{\Omega_0},\quad T/\Omega_{0}\ll1.     
\end{equation}

\begin{figure}[t]
\centering\includegraphics[width=.5\columnwidth]{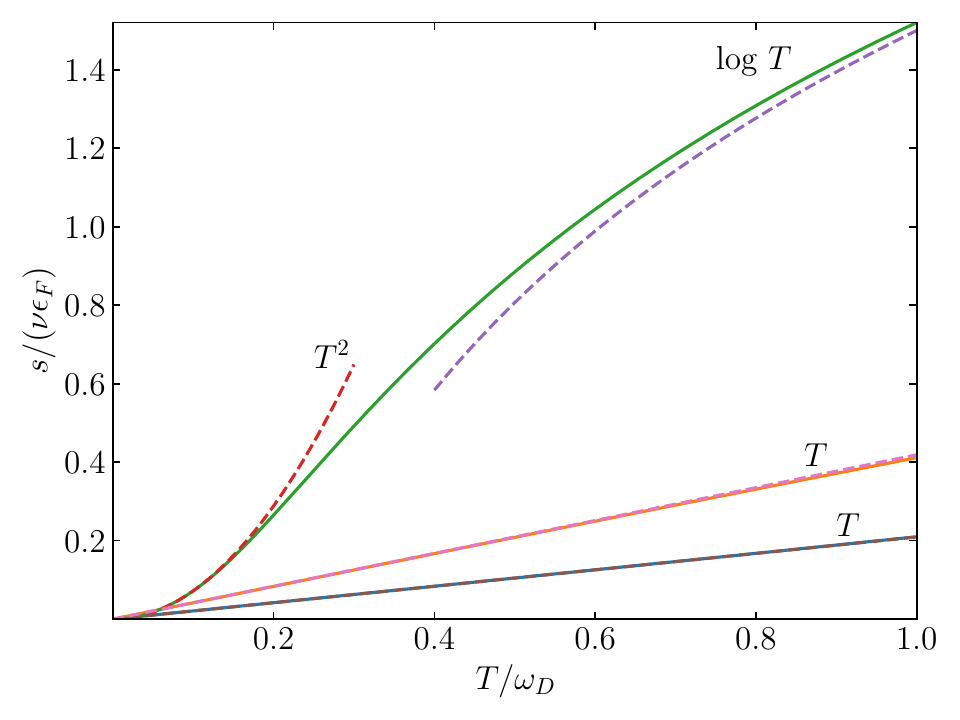}
\caption{Entropy density in the impurity regime in units of $\nu\varepsilon_{F}$. Green, blue, and orange lines are the bosonic, electronic, and interacting ($\mathcal{G}$-$\Sigma$) contributions, respectively. Dashed lines denote the asymptotic behavior in the high and low-temperature limits. The energy scales are $\omega_D=M_0=\varepsilon_{F}=\Lambda/2=\Omega_0/20$, corresponding to $\lambda=15\pi/16$.}
\label{fig:EntropyIM}
\end{figure}

As for the bosonic contribution, we may utilize the fact that both the Landau damping term and the renormalized mass are suppressed at strong coupling and low temperatures ($\Pi(i\Omega_{m\neq0})\propto 1/\lambda$ and $M^2\sim e^{-\lambda}$). Thus, the bosonic contribution to the entropy density is given by Eq.~\eqref{eq:boson-free-entropy}, with $\omega_B=\omega_D$ for $T\ll \Omega_0$. In the impurity regime, the energy scale for the electrons is $\Omega_0$, where $\Omega_0\propto\lambda$. Therefore, in the strong-coupling limit, $\Omega_0 \gg T$ and the electrons are `cold'~\cite{EsterlisSchmalian2019}. It is therefore unsurprising that the fermionic contribution to the entropy density $s_{F}\propto1/\lambda\rightarrow0$ in the strong-coupling limit. Thus, the bosonic contribution to the entropy density is dominant at intermediate temperatures and strong coupling. The entropy-density for the finite bandwidth case with strong coupling is displayed in Fig.~\ref{fig:EntropyIM}.

\section{Fermionic self-energy}

\subsection{Fermionic self-energy in the infinite-bandwidth limit}

\subsubsection*{Zero-temperature analysis}

At sufficiently weak coupling the fermionic self-energy near $T_c$ is well approximated by its zero-temperature limit. The fermionic self-energy on the imaginary axis at zero temperature is given by 
\begin{align}
\Sigma(i\omega) &= g^2\int_{-\infty}^{\infty}\mathcal{G}(i\omega^{\prime})\mathcal{D}(i\omega-i\omega^{\prime})\frac{d\omega^{\prime}}{2\pi} \nonumber \\
&= -i \frac{\lambda}{2}\int_{-\infty}^{\infty}L(i\omega-i\omega^{\prime})\text{sgn}(\omega^{\prime})d\omega^{\prime} \nonumber \\
&= -i \lambda\;\text{sgn}(\omega)\int_{0}^{|\omega|}L(i\omega^{\prime})d\omega^{\prime}. 
\end{align}
Here we have used the fact that the effective bosonic propagator $L(i\Omega)$ is an even function of $\Omega$ and has a limit of zero as $|\Omega|\rightarrow\infty$. For $|\omega|/\omega_D\ll 2\pi\lambda_E$ and a quantum-critical boson, the Landau damping term in the effective bosonic propagator is dominant. In this case, the fermionic self-energy is then
\begin{align}\nonumber \label{eq:MFL-SE1}
    \Sigma(i\omega)&=-i \lambda\;\text{sgn}(\omega)\int_{0}^{|\omega|}\log\left(1+\frac{\varepsilon_{F}}{2\pi \lambda |\omega^{\prime}|}\right)d\omega^{\prime}\\
    &\simeq-i\lambda\omega\;\log\left(\frac{e\varepsilon_{F}}{2\pi\lambda|\omega|}\right); \quad  |\omega|/\omega_{D} \ll \frac{1}{2\pi\lambda_{E}}, 2\pi\lambda_{E}.
\end{align}
Still within the weak-coupling regime (where $T_c$ is exponentially suppressed), the boson may be underdamped (away from $\omega=0$). This is the case when $|\omega|/\omega_D \gg 2\pi\lambda_E$. Computing the fermionic self-energy at zero temperature for a quantum-critical boson in such a regime, we find 
\begin{align}\nonumber \label{eq:MFL-SE2}
    \Sigma(i\omega)&=-i \lambda\;\text{sgn}(\omega)\int_{0}^{|\omega|}\log\left(1+\left(\frac{\omega_D}{\omega^{\prime}}\right)^2\right)d\omega^{\prime}\\
     &\simeq -2i\lambda\omega\; \log\left(\frac{e\omega_D}{|\omega|}\right); \quad 2\pi\lambda_{E} \ll |\omega|/\omega_{D} \ll 1.
\end{align}
Both regimes display marginal-Fermi-Liquid physics~\cite{Varma1989}, although with different quasi-particle lifetimes and UV cutoffs for the MFL form. In particular, the overdamped regime has a quasi-particle lifetime $\tau =1/(\pi\lambda|\omega|)$, while the underdamped regime has $\tau =1/(2\pi\lambda|\omega|)$ for $\omega\ll \omega_D$.

\subsubsection*{Exact result for non-zero temperatures}

In the infinite bandwidth limit, the fermionic self-energy is
\begin{equation}
\Sigma\left(i \omega_n\right)=-i \pi\nu g^2T \sum_{\omega_{n^\prime}} \operatorname{sgn}\left(\omega_{n^{\prime}}\right) \mathcal{D}\left(i \omega_n-i \omega_{n^{\prime}}\right).
\label{eq:SigmaSum1}
\end{equation}
By isolating the resonant term ($\omega_n=\omega_{n^\prime}$), we can express the self-energy as
\begin{align}
\Sigma\left(i \omega_n\right) &= -i \pi\nu g^2T \left[\text{sgn}(\omega_n)\mathcal{D}\left(0\right)+\sum_{\omega_{n^\prime}\neq\omega_n} \operatorname{sgn}\left(\omega_{n^{\prime}}\right) \mathcal{D}\left(i \omega_n-i \omega_{n^{\prime}}\right) \right] \nonumber\\
&= -i \pi\nu g^2  T\, \mathrm{sgn}(\omega_n) \left[\mathcal{D}(0)+2\sum_{0 < \Omega_m<|\omega_n|} \mathcal{D}(i\Omega_m)\right].
\label{eq:SigmaSum2}
\end{align}
The exact expression for the fermionic self-energy can be obtained by performing the Matsubara frequency summation. To do this, we use the digamma function $\psi(z)$, which is defined by $\psi(z)=(d/dz)\log\Gamma(z)$, where $\Gamma(z)$ denotes the Gamma function~\cite{WhittakerWatsonBook}. Note that $\Gamma(z)$ is a holomorphic single-valued complex function of a complex variable for all $z\in\mathbb{C}$ excluding non-positive integers. 

\begin{figure}[ht]
\centering\includegraphics[width=1\columnwidth]{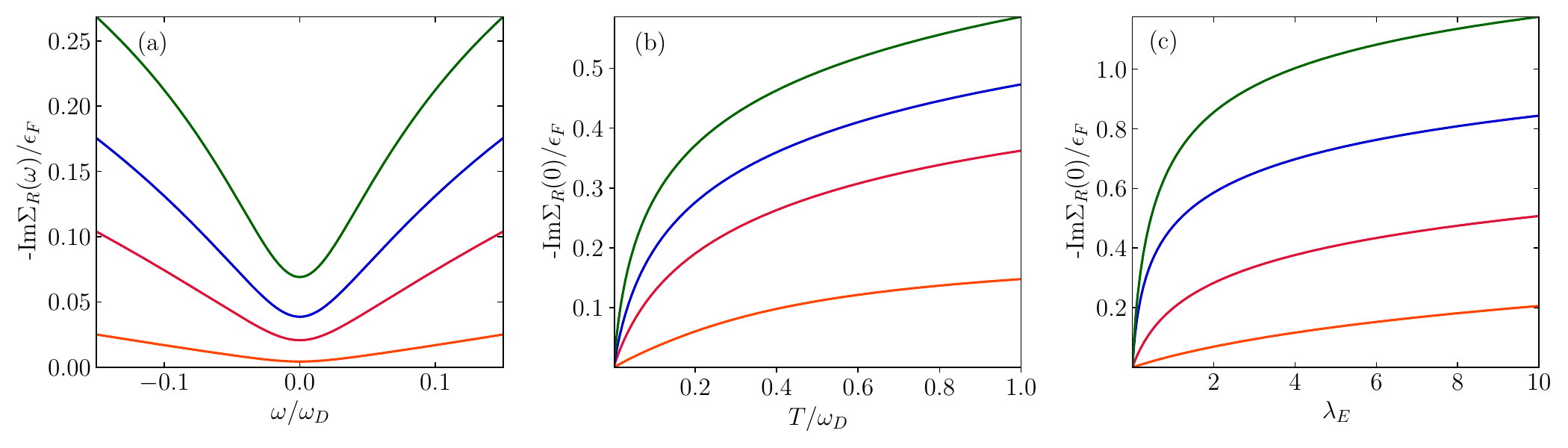}
\caption{Imaginary part of the Fermionic self-energy in units of $\varepsilon_{F}$ in the infinite bandwidth limit, computed using Eq.~\eqref{eq:Exact_elec_SE}. The fermionic self-energy is given in (a) as a function of frequency (units $\omega_D$) evaluated at $T=\omega_D/100$, (b) a function of temperature (units $\omega_D$), (c) as a function of dimensionless coupling strength, $\lambda_E$. For (a) and (b), $\lambda_E = 0.1,0.5,1,2$ for the orange, red, blue, and green curves respectively. (b) and (c) are evaluated at zero frequency, $\omega=0$. For (c), $T/\omega_D = 1/100, 1/10, 1, 10$ for the orange, red, blue, and green curves respectively.}
\label{fig:Eval_SE}
\end{figure}

Since $\Sigma(i\omega_{n})$ is an odd function of Matsubara frequency, we can obtain its value for positive $\omega_{n}$ and then deduce it on the negative Matsubara axis by symmetry. Therefore, let us assume that $n\geq0$ for convenience. 
Interchanging the momentum integration and Matsubara frequency summation (which is permissible in this case), we obtain 
\begin{align}
\Sigma(i\omega_{n})	&= -i\pi\lambda T\left[L(0)+8\pi c^{2}\sum_{k=1}^{n}\mathcal{D}(i\Omega_{k})\right] \nonumber\\
&=	-i\pi\lambda T\left\{ L(0)+2\int_{0}^{\omega_{D}^{2}}\left[\sum_{k=1}^{n}\frac{1}{\Omega_{k}^{2}+y+M^{2}+2\pi\nu^{2}g^{2}|\Omega_{k}|}\right]dy\right\}.
\end{align}
The summation over $k$ can be performed as follows. Let $k_{+}$ and $k_{-}$ be the roots of the equation $Ak^{2}+Bk+C=0$, where $A=(2\pi T)^{2},B=(2\pi\nu g)^{2}T$, and $C=y+M^{2}$; $T>0$ and we assume that $k_{\pm}$ are not equal and are not integers. 
Thus,  
\begin{align}
\sum_{k=1}^{n}\frac{1}{Ak^{2}+Bk+C} &= \frac{1}{A(k_{+}-k_{-})}\sum_{k=1}^{n}\left[\frac{1}{k-k_{+}}-\frac{1}{k-k_{-}}\right] \nonumber\\
&= \frac{1}{A(k_{+}-k_{-})}\left[\psi(n+1-k_{+})-\psi(1-k_{+})-\psi(n+1-k_{-})+\psi(1-k_{-})\right].
\end{align}
In the last step, we obtained the sum of the series by using the recurrence relation for the digamma function: $\psi(z+1)=\psi(z)+1/z$. Thus, the self energy becomes 
\begin{align}
\Sigma(i\omega_{n})	&= -i\pi\lambda TL(0)-2i\pi\lambda T\int_{M^{2}}^{\omega_{D}^{2}+M^{2}}\left[\frac{\psi(n+1-k_{+})-\psi(1-k_{+})-\psi(n+1-k_{-})+\psi(1-k_{-})}{2(k_{+}-k_{-})}\right]dC \nonumber\\
&= -i\pi\lambda TL(0)-2i\pi\lambda T\left.\log\left[\frac{\Gamma(n+1-k_{+})\Gamma(n+1-k_{-})}{\Gamma(1-k_{+})\Gamma(1-k_{-})}\right]\right|_{C=M^{2}}^{C=M^{2}+\omega_{D}^{2}}.
\end{align}
If we introduce the Pochhammer symbol~\cite{WhittakerWatsonBook} defined by $(x)_n\equiv\Gamma(x+n)/\Gamma(x)$, and define the variables $t\equiv T/\omega_{D}, \bar{m}\equiv M/\omega_D$, and the parameter $\lambda_E=g^2\nu^2/\omega_D$, then the exact result for the fermionic self energy is given by 
\begin{equation}
\frac{\Sigma(i\omega_{n})}{\varepsilon_{F}}=-i\pi\lambda_{E}t\left\{ L(0)+2\log\left[\frac{\left(1+\dfrac{\lambda_{E}-\sqrt{\lambda_{E}^{2}-(\bar{m}^{2}+1)/\pi^{2}}}{2t}\right)_{n}}{\left(1+\dfrac{\lambda_{E}+\sqrt{\lambda_{E}^{2}-\bar{m}^{2}/\pi^{2}}}{2t}\right)_{n}}\frac{\left(1+\dfrac{\lambda_{E}+\sqrt{\lambda_{E}^{2}-(\bar{m}^{2}+1)/\pi^{2}}}{2 t}\right)_{n}}{\left(1+\dfrac{\lambda_{E}-\sqrt{\lambda_{E}^{2}-\bar{m}^{2}/\pi^{2}}}{2t}\right)_{n}}\right]\right\}.
\label{eq:Exact_elec_SE}
\end{equation}
The imaginary part of the retarded fermionic self-energy in the infinite bandwidth limit is evaluated (using the renormalized boson mass determined in Sec.~\ref{sec:RenormalizedBosonMass}) in Fig.~\ref{fig:Eval_SE}. 

\subsubsection*{Real frequency axis}

In this section, we formulate the normal-state equations for the (2+1)d Y-SYK model on the real frequency axis, following the analogous prescription~\cite{Marsiglio1988} for Eliashberg theory. The advantage of this formulation compared to the conventional approach~\cite{AGD, Karakozov1975, Schmalian1996, Esterlis2025} is that a hybrid set of real and imaginary frequency axes equations are obtained, which allows one to easily recover the familiar imaginary frequency axis equations. If we let $z=\omega+i0^{+}$, then the real frequency axis Green's functions for the Y-SYK model (in the infinite bandwidth limit) are
\begin{align}
\mathcal{G}_{R}(z) &= -i\nu\pi, \\
\mathcal{D}_{R}(z) &= \frac{1}{4\pi c^{2}}\log\left[1-\frac{\omega_{D}^{2}}{z^{2}-M_0^2+\Pi_{R}(z)}\right], \\
\Sigma_{R}(z) &= -i\nu\pi g^{2}T\sum_{m=-\infty}^{\infty}\textrm{sgn}(\omega_{m})\mathcal{D}(z-i\omega_{m})  \nonumber \\
&\quad -i\nu\pi g^{2}\int\text{Im}\mathcal{D}_{R}(y)\left[\coth\left(\frac{\beta y}{2}\right)-\tanh\left(\frac{\beta}{2}(y-z)\right)\right]\frac{dy}{2\pi}, \\
\Pi_{R}(z) &= -2g^{2}\int\left[\tanh\left(\frac{\beta}{2}(y-z)\right)-\tanh\left(\frac{\beta y}{2}\right)\right]\text{Im}\mathcal{G}_{R}(y)\mathcal{G}_{R}(y-z)\frac{dy}{2\pi}.
\end{align}
The above equations are analogous to those in Ref.~\cite{Marsiglio1988}. The distinction here is that there is also a bosonic self energy. The real part of $\mathcal{G}_{R}$ is zero and thus so too is the real part of $\Pi_{R}$. As a check on the above equations, we directly compute $\text{Im}\Pi_{R}$ and compare the result with analytical continuation of the imaginary frequency axis expression. 
\begin{align}
\textrm{Im}\Pi_{R}(z) &= -2g^{2}\int_{-\infty}^{\infty}\left[\tanh\left(\frac{\beta}{2}(y-z)\right)-\tanh\left(\frac{\beta y}{2}\right)\right] \text{Im}\mathcal{G}_{R}(y)\textrm{Im}\mathcal{G}_{R}(y-z)\frac{dy}{2\pi} \nonumber\\
 &= -2(\nu\pi g)^{2}\int_{-\infty}^{\infty}\left[\tanh\left(\frac{\beta}{2}(y-z)\right)-\tanh\left(\frac{\beta y}{2}\right)\right]\frac{dy}{2\pi} \nonumber\\
&= 2\pi\nu^{2}g^{2}\Omega.
\end{align}
Thus, the boson self energy, on the real frequency axis, is $\Pi_{R}(\Omega+i\delta)=2\pi i\nu^{2}g^{2}\Omega$. This result agrees with the analytical continuation of the normal-state result on the Matsubara frequency axis, namely, $\Pi(i\Omega_{m})=-2\pi\nu^{2}g^{2}|\Omega_{m}|$. Using the form of the boson propagator in Eq.~\eqref{eq:BosonL}, the imaginary part of the fermionic self energy is given by
\begin{equation}
\Sigma_R^{\prime\prime}(\omega) = -\frac{\lambda}{2}\int_{-\infty}^{\infty} \text{Im}\log \left[1-\frac{1}{\left(\Omega/\omega_D\right)^2-\left(M/\omega_D\right)^2+2 \pi i \lambda \Omega/\varepsilon_F}\right]\left[\text{coth}\left(\frac{\Omega}{2T}\right)+\text{tanh}\left(\frac{\omega-\Omega}{2T}\right)\right]d\Omega.
\label{eq:Real-axis-self-energy}
\end{equation}
Equation~\eqref{eq:Real-axis-self-energy} agrees with the analytical continuation of Eq.~\eqref{eq:Exact_elec_SE}, appearing in Fig.~\ref{fig:Eval_SE}.

\subsection{Fermionic self-energy in the finite-bandwidth limit}

\subsubsection*{Spectral function}

In the finite-bandwidth limit defined by the separation of scales $|i\omega_n-\Sigma(i\omega_n)| \gg \Lambda/2$, the 2d Y-SYK theory admits an NFL impurity-like solution~\cite{EsterlisSchmalian2019} for the fermionic self-energy. In the strong-coupling limit, where this regime occurs, the self-energy is given by Eq.~\eqref{eq:impurity_self-energy}, and the inverse quasi-particle lifetime is $1/(2\tau) \sim \Omega_0/2= 16 \lambda \Lambda/(3\pi \kappa)$. In the strong coupling limit, $\Omega_0$ becomes the largest scale: $\Omega_0 \gg \Lambda$, thus describing a bad metal.

\begin{figure}[ht]
\centering\includegraphics[width=.5\columnwidth]{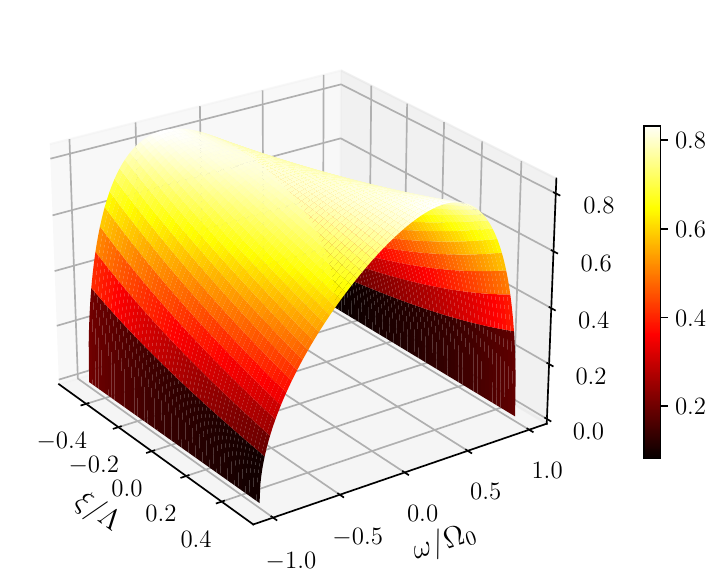}
\caption{The fermionic spectral function, $A(\omega,\xi)$, in units of $\Lambda^{-1}$, as a function of frequency $\omega/\Omega_0$ and isotropic dispersion energy $\xi/\Lambda$ evaluated for $\Omega_0/\Lambda=5$.}
\label{fig:spectral-function}
\end{figure}

By analytically continuing Eq.~\eqref{eq:impurity_self-energy}, we obtain the real-axis fermionic self-energy within the impurity regime
\begin{subequations}
\begin{align}
\Sigma_R^{\prime}(\omega) &= \frac{1}{2}
\begin{cases}
        \quad\quad\quad \omega, \quad &|\omega|<\Omega_0\\
        \omega-\text{sgn}(\omega)\sqrt{\omega^2-\Omega_0^2}, &|\omega|>\Omega_0\\
\end{cases}. \\
\Sigma_R^{\prime\prime}(\omega) &= \frac{1}{2}
\begin{cases}
        -\sqrt{\Omega_0^2-\omega^2},\quad &|\omega|<\Omega_0\\
        \quad\quad\quad0, \quad &|\omega|>\Omega_0 \\
\end{cases}. 
\end{align}
\end{subequations}
These functions obey the Kramers-Kronig relations (and of course reduce to the self energy on the Matsubara axis), and thus they are the unique analytical continuation of the self energy. The imaginary part of the self energy exhibits a sharp transition from nonzero to zero imaginary part as the frequency approaches $\Omega_0$. The UV regime can be better understood by evaluating the spectral function, $A(\omega)\equiv-2\text{Im}G_R(\omega)$. For $|\omega|>\Omega_0$, $A(\omega,\xi) = 2\pi \delta(\omega-\xi-\text{sgn}(\omega)(|\omega|-\sqrt{\omega^2-\Omega_0^2})/2)$. Since $-\Lambda/2\leq \xi \leq \Lambda/2$, there are no solutions to $\xi = \omega-\text{sgn}(\omega)(|\omega|-\sqrt{\omega^2-\Omega_0^2})/2$ for $\Lambda<\Omega_0$. Thus, the full spectral function in the impurity regime is given by
\begin{equation}
\label{eq:FBW-spectral-function}
A(\omega,\xi) = \frac{4\sqrt{\Omega_0^2-\omega^2}}{(2\xi)^2-4\omega \xi + \Omega_0^2}\;\Theta(\Omega_0-|\omega|).
\end{equation}
Evidently, the $|\omega|>\Omega_0$ region of frequencies does not contribute to the electromagnetic transport within the impurity regime, even at high temperatures. In Fig.~\ref{fig:spectral-function} the spectral function is plotted.

\subsubsection*{Bad-metal-to-strange-metal crossover}
\label{sec:crossover}

The infinite-bandwidth analysis utilizes a Lagrange-multiplier constraint to ensure that the renormalized boson mass vanishes at zero temperature, as employed in Ref.~\cite{Esterlis2021}. In contrast, the finite-bandwidth analysis does not utilize a constraint to enforce quantum criticality. Instead, we observe that the impurity regime exhibits self-tuning to the QCP, as found in the (0+1)d Y-SYK model considered in Refs.~\cite{EsterlisSchmalian2019, Wang2020}. With a self-consistent evaluation of the renormalized boson mass in a finite-bandwidth scenario, we find the renormalized boson mass in the (2+1)d Y-SYK model to vanish at zero temperature for an arbitrary-valued (and temperature-independent) bare bosonic mass $M_0$. The crossover between the marginal-Fermi-liquid-like to impurity-like electronic self-energy is driven by the temperature and coupling-strength scaling of the renormalized boson mass, and can be understood by examining the quantum and thermal contributions to the electronic self-energy.

The electronic self-energy can be split into its quantum (non-resonant $n\neq n^\prime$) and thermal (resonant $n=n^\prime$) contributions. At low temperatures, $T\ll \omega_c$, the quantum contribution is of the form of a marginal Fermi liquid. In the infinite-bandwidth limit (while incorporating a constraint term in the action), the thermal contribution is zero at zero temperature and nonzero for nonzero temperatures. For a finite bandwidth (self-tuned to the QCP), the thermal contribution is non zero, even at zero temperature. By increasing the coupling in a finite-bandwidth scenario, the thermal contribution becomes dominant and the system crosses over from a marginal Fermi liquid to a bad metal.

Although the transition from marginal Fermi liquid to bad metal is driven by the \emph{thermal} contribution to the electronic self-energy, the transition occurs at arbitrarily low temperatures for sufficiently strong coupling. Within the impurity regime, at low temperatures ($T\ll\Omega_0$) $\Omega_{0}$ scales as $\Omega_0\propto\lambda$, as in Eq.~\eqref{eq:Omega0}. Consequently, the smaller the renormalized boson mass, the larger the impurity energy scale will be, which in turn ensures the smallness of the renormalized boson mass. In this sense, a dominant thermal contribution to the fermionic self-energy is concomitant with being near the quantum critical point, where $M(T=0)=0$, in a strong-coupling finite-bandwidth scenario.

\section{Effects of spatially random mass disorder}

We consider quenched~\cite{Dotsenko} random-mass disorder in the fermion mass~\cite{Patel2023}
\begin{equation}
S_v=\frac{1}{\sqrt{N}} \sum_{\sigma=\uparrow\downarrow}\sum_{i, j=1}^N \int v_{ij}({\bf r}) \, \psi_{i\sigma}^\dagger(\tau,{\bf r}) \psi_{j\sigma}(\tau,{\bf r})\ d\tau d^{2}r.
\end{equation}
Here, the spatially dependent mass $v_{ij}({\bf r})$ is sampled from a Gaussian orthogonal ensemble, local in real space with zero mean and non-zero variance: $\langle v_{ij}(\bm{r})\rangle=0, \ \langle v_{ij}(\bm{r}) v_{ab}(\bm{r}^{\prime})\rangle = 2v^2\delta({\bf r}-{\bf r}')\delta_{ia}\delta_{jb}$. The self energy is now replaced by $\Sigma(i\omega_n)\rightarrow\Sigma(i\omega_n)+v^2\mathcal{G}(i\omega_n)$ in the momentum-space saddle-point Eq.~\eqref{eq:Sigma_saddlept} for the fermionic self-energy.

\subsection{Infinite-bandwidth limit}

In the infinite bandwidth limit, $\mathcal{G}(i\omega_n)=-i\pi\nu\;\text{sgn}(\omega_n)$, and thus spatially random disorder gives the additional contribution $-i\pi v^2\nu\;\text{sgn}(\omega_n)$ to the fermionic self-energy. On the real axis, spatially random potential disorder within the infinite bandwidth limit provides a temperature independent elastic scattering rate $\Gamma_0\equiv \pi v^2\nu$, which uniformly broadens the electronic spectral function. Finite broadening (nonzero inverse single-particle lifetime) ensures that both the normal-state DC electrical conductivity and shear viscosity are finite. In the infinite bandwidth limit ($\Lambda/2\gg|i\omega_n-\Sigma(i\omega_n)|$), the form of the momentum integrated electronic Green's function is independent of the spatially random potential disorder. As a result, the bosonic self-energy and propagator, as well as the total entropy density, are independent of the fermionic potential disorder.

\subsection{Finite-bandwidth limit}
\subsubsection*{Robustness to potential disorder}

To leading order in the electronic bandwidth $\Lambda$, $\mathcal{G}(i\omega_n)=\nu\Lambda/(i\omega_n-\Sigma(i\omega_n))$. In the impurity regime, the resonant ($n=n^{\prime}$) contribution to the Y-SYK interaction is dominant
\begin{equation}
\Sigma(i\omega_n) = \nu \Lambda \left[\frac{v^2}{i\omega_n-\Sigma(i\omega_n)}+g^2T\sum_{\omega_{n^{\prime}}}\frac{\mathcal{D}(i\omega_n-i\omega_{n^{\prime}})}{i\omega_n^{\prime}-\Sigma(i\omega_n^{\prime})}\right] \approx\nu \Lambda \left(\frac{v^2 + g^2 T \mathcal{D}(0)}{i\omega_n-\Sigma(i\omega_n)}\right).
\end{equation}
The impurity scattering rate is thus renormalized by $\Omega_0\rightarrow\widetilde{\Omega}_0\equiv2\sqrt{\Lambda(\nu v^2+\lambda T L(0))}$. As a result, the low-temperature self-consistent equation for the renormalized boson mass is given by $M^2=M_0^2 -16(g\nu\Lambda)^2/(3\pi\widetilde{\Omega}_0)$. Let us define the parameter $\theta\equiv\lambda\Lambda/(3\pi\kappa)^2$, such that the renormalized boson mass is given by $M\sim\omega_D\text{exp}(-\theta/(2T))$ for $T\ll\theta$ in the absence of potential disorder.  Including potential disorder, the parameter $\theta$ is modified. In the impurity regime, and for low temperatures $\theta\rightarrow\widetilde{\theta}\approx\theta-v^2\nu/\lambda $. 

For $\widetilde{\theta}>0$, $\underset{T\rightarrow0}{\lim}M(T)=0$, demonstrating the Y-SYK interaction maintains its feature of ``self-tuning'' to the QCP within the impurity regime. In which case, for $T\ll\widetilde{\theta}$, the renormalized boson mass is given by $M\sim\omega_D \text{exp}(-\widetilde{\theta}/(2T))$. However, for $\widetilde{\theta}<0$, $\underset{T\rightarrow0}{\lim}M(T)\neq0$, and the system is taken away from the QCP. We observe both the self-tuning quantum-critical feature and the energy scale at which Fermionic excitations scatter to be resilient to impurities for $T\ll\widetilde{\theta}$, provided the Y-SYK interaction is dominant over the potential disorder. Above a threshold $v>v_*$, defined
\begin{equation}
v_*\equiv  \frac{16 \lambda}{3\pi \kappa}\sqrt{\frac{\Lambda}{\nu}},
\end{equation}
we find the renormalized bosonic mass to be detuned from the QCP. For $v<v_*$ and $T\ll\widetilde{\theta}$, $L(0)\sim \widetilde{\theta}/T$, and thus 
\begin{equation}
\widetilde{\Omega}_0 \sim 2\sqrt{\Lambda(\nu v^2 + \lambda\widetilde{\theta})} = 2\sqrt{\lambda\Lambda \theta} = \Omega_0.
\end{equation}
Demonstrating that at the QCP ($v<v_*$), the Y-SYK electron boson interaction leads to an impurity energy scale $\Omega_0$, unaffected by potential disorder in the limit $T/\widetilde{\theta}\rightarrow0$ in the strong coupling regime.

\subsubsection*{Violation of Matthiessen’s rule}

The robustness to potential disorder of the impurity regime in the $T/\tilde{\theta}\rightarrow0$ limit may be directly inferred from the condition of quantum criticality. For $T\ll\widetilde{\Omega}_0$, the self-consistent equation for the renormalized boson mass is given by $M^2= M_0^2-16(g\nu\Lambda)^2/(3\pi\widetilde{\Omega}_0)$. In the vicinity of the QCP, $M/M_0\ll1$, and the self-consistent equation for the renormalized boson mass places the constraint $\widetilde{\Omega}_0\approx16(g\nu\Lambda)^2/(3\pi M_0^2)\equiv\Omega_0$ on the fermionic scale. Quantum-critical solutions within the impurity regime are thus constrained in a manner independent of the potential disorder.

This mechanism corresponds to a tradeoff between electron-boson and electron-impurity scattering. Electron-impurity scattering increases the renormalized boson mass, requiring electronic excitations to have a greater energy to excite the bosons. This results in reduced electron-boson scattering so that the overall scattering rate is fixed. This mechanism occurs until the impurity regime breaks down, either by a violation of the assumed separation of scales or by the impurity scattering rate exceeding the electron-boson scattering rate ($v>v_*$), destroying the QCP. In summary, within the impurity regime and in the vicinity of the QCP ($T\ll\widetilde{\theta}$):
\begin{equation}
\Sigma=\Sigma_{\text{Y-SYK}}(v)+\Sigma_{\text{Disorder}}(v)\sim\Sigma_{\text{Y-SYK}}(0);\quad M(v)\sim \text{exp}(v^2\nu/(2\lambda T))M(0).
\end{equation}

Matthiessen’s rule~\cite{ZimanBook} states that the transport scattering rate is comprised of the contributions from independent scattering channels (e.g. electron-impurity, electron-phonon) which add linearly to give the resistivity. We find that the Y-SYK and potential disorder contributions are not independent within the impurity regime. In fact, for $T\ll\widetilde{\theta}$, the total scattering rate is independent of the potential disorder, demonstrating a strong violation of Matthiessen’s rule. Revealing this mechanism within the bad-metal regime is significant, as it has been suggested that the violation of Matthiessen’s rule is a key signature of the high-temperature regime of unconventional superconductors~\cite{Liu2024}.

\section{Electrical conductivity}

\subsection{Kubo formula and spectral representation}
 
The real part of the longitudinal DC electrical conductivity is defined by the following Kubo formula:
\begin{align} \nonumber
    \sigma &= -\lim_{\Omega \rightarrow 0}\frac{1}{\Omega}\text{Im}K^{xx}_R(\Omega,0),\\
    K^{xx}_R(\Omega,0) &= i \int_{0}^{+\infty} d t \int e^{i \Omega t} \left\langle\left[\hat{J}_{x}(\bm{r}, t), \hat{J}_{x}(\bm{0}, 0)\right]\right\rangle d^2r.
\end{align}
The components of the current operator are given by $\hat{J}_x$. Using the Matsubara formalism, we express the Fourier transform of the time-ordered correlation function as
\begin{equation}
K^{xx}(i\Omega_m,0) = 2e^2T\sum_{\omega_n}\int \left(\frac{k_x}{m}\right)^2 G(i\omega_n, \bm{k}) G(i\omega_n+i\Omega_m, \bm{k})\frac{d^2k}{(2\pi)^2}.
\end{equation}
Using the spectral representation $G(z) = \int_{\mathbb{R}} \frac{A(\omega)}{\omega-z}\frac{d\omega}{2\pi}$, where $\text{Im}z\neq0$ and $A(\omega) = -2 \text{Im}G_R(\omega)$, we obtain
\begin{equation}
K^{xx}(i\Omega_m,0) = 2e^2T\sum_{\omega_n}\int \biggl[\int \left(\frac{k_x}{m}\right)^2 \frac{A(\omega, \bm{k})}{i \omega_n -\omega}\frac{A(\omega^\prime , \bm{k})}{i \omega_n + i \Omega_m - \omega^\prime}\frac{d^2 k}{(2\pi)^2}\biggr] \frac{d \omega d \omega^\prime}{(2\pi)^2}.
\end{equation}
Evaluating the Matsubara sum then performing the analytical continuation of the external frequency $i\Omega_m \rightarrow \Omega+i0^+$ leads to
\begin{equation}
K^{xx}_R(\Omega,0) = 2e^2\int \left[\int \left(\frac{k_x}{m}\right)^2 A(\omega, \bm{k})A(\omega^\prime, \bm{k})\frac{n_F(\omega)-n_F(\omega^\prime )}{\omega-\omega^\prime+\Omega+i0^+}\frac{d^2 k}{(2\pi)^2}\right]\frac{d \omega d \omega^\prime}{(2\pi)^2}.
\end{equation}
Using the Sokhotski–Plemelj theorem and the real-valuedness of the spectral function, we obtain
\begin{align}\nonumber
    \text{Im}K^{xx}_R(\Omega,0) &=-\int \frac{d \omega d \omega^\prime}{2\pi}\int \left(\frac{k_x}{m}\right)^2 A(\omega, \bm{k})A(\omega^\prime, \bm{k})\left(n_F(\omega)-n_F(\omega^\prime)\right)\delta(\omega-\omega^\prime+\Omega) \frac{d^2 k}{(2\pi)^2}\\ 
    &=-\int \frac{d \omega}{2\pi}\int \left(\frac{k_x}{m}\right)^2 A(\omega, \bm{k})A(\omega+\Omega, \bm{k})\left(n_F(\omega)-n_F(\omega + \Omega)\right)\frac{d^2 k}{(2\pi)^2}.
\end{align}
We use the spectral representation to express the longitudinal DC conductivity as
\begin{equation}
\sigma =e^2\bigintssss \left[\bigintssss \left(\frac{k_x}{m}\right)^2 A^2(\omega, \bm{k}) \frac{d^2 k}{(2\pi)^2}\right]\left(-\frac{\partial n_F}{\partial\omega}\right)\frac{d \omega}{2\pi} .
\label{eq:conductivity-intermediate}
\end{equation}
Note that the Fermi-Dirac distribution function depends only on frequency $\omega$.  
The Green's function takes the form
\begin{equation}
G_R(\omega, \bm{k}) = (\omega-\xi_{\bm{k}} - \Sigma_R(\omega))^{-1}.
\label{eq:GreensFn}
\end{equation}
The (retarded) self energy is a complex function of frequency and can be written in terms of its real and imaginary parts as $\Sigma_R(\omega) = \Sigma^\prime_R(\omega) + i \Sigma^{\prime\prime}_R(\omega)$. The spectral function then evaluates to
\begin{equation}
A(\omega, \bm{k}) = \frac{2|\Sigma^{\prime\prime}_R(\omega)|}{(\omega-\xi_{\bm{k}}-\Sigma^{\prime}_R(\omega))^2+(\Sigma^{\prime\prime}_R(\omega))^2}.
\label{eq:SpectralFn}
\end{equation}
For an isotropic dispersion relation $\xi_{\bm{k}}=k^2/(2m)-\Lambda/2$ with bandwidth $\xi_{\bm{k}}\in[-\Lambda/2,\Lambda/2]$, and for an arbitrary function $F(k)$, we have
\begin{equation}
\int\left(\frac{k_x}{m}\right)^2F(k)\frac{d^2k}{(2\pi)^2} = \frac{1}{(2\pi m)^2}\int_{k_{\textrm{min}}}^{k_{\textrm{max}}}  k^3F(k)\ dk\int_{0}^{2\pi}\cos^2\theta\  d\theta = \int_{-\Lambda/2}^{\Lambda/2}\left(\xi+\Lambda/2\right)F(\xi)\ \frac{d\xi}{2\pi}.
\label{eq:AngularInt}
\end{equation}
The DC conductivity for the isotropic dispersion with symmetric bandwidth is
\begin{equation}
\sigma =e^2\bigintssss_{-\infty}^{\infty} \left[\bigintssss_{-\Lambda/2}^{\Lambda/2}  \left(\xi+\Lambda/2\right)A^2(\omega,\xi)\ \frac{d\xi}{2\pi}\right] \left(-\frac{\partial n_F}{\partial\omega}\right) \frac{d\omega}{2\pi}.
\label{eq:conductivity-arb-bw}
\end{equation}

The relevant transport regimes can be characterized in terms of the following energy scales: temperature $T$, electronic bandwidth $\Lambda$, and spectral width $\Gamma(\omega)\equiv |\Sigma_R^{\prime\prime}(\omega)|\equiv1/(2\tau(\omega))$. The zero-frequency spectral width is $\Gamma=|\Sigma_R^{\prime\prime}(0)|$. Here, it is assumed that $\Gamma(\omega)$ is described solely by momentum-relaxing scattering.  We consider conductivity in the infinite-bandwidth and finite-bandwidth limits, with the latter regime describing bad-metal (BM) transport at low to intermediate temperatures and thermal-incoherent-metal (TIM) transport at ultra-high temperatures. The electronic dispersion is assumed to be quadratic and isotropic and is evaluated in the continuum limit. All formulae are given in two spatial dimensions but can be straightforwardly generalized to arbitrary spatial dimensions.

\subsection{Limiting formulae}

\subsubsection*{Low temperatures -- Arbitrary bandwidth}

In the case where temperature is the smallest energy scale, $T\ll\Lambda,\Gamma(\omega)$, the derivative of the Fermi-Dirac distribution function can be approximated as $n_{F}^\prime(\omega) \approx -\delta(\omega)$, and conductivity becomes
\begin{equation}
\sigma=\frac{e^2}{2\pi}\int_{-\Lambda/2}^{\Lambda/2}(\xi+\Lambda/2)A^2(0,\xi)\frac{ d\xi}{2\pi}. 
\end{equation}
The zero-frequency spectral function is $A(0,\xi)=2|\Sigma_R^{\prime\prime}(0)|/[(\xi+\Sigma_R^{\prime}(0))^2+ \Sigma_R^{\prime\prime}(0)^2]$. Assuming $\Sigma_R^{\prime}(0)\ll\Sigma_R^{\prime\prime}(0),\Lambda$, the low-temperature conductivity is given by 
\begin{equation}\label{eq:cond-arb-bw}
\sigma  = \left(\frac{e^2\Lambda}{4\pi \Gamma}\right)\frac{2}{\pi}\left[\text{arctan}\left(\frac{\Lambda}{2\Gamma}\right) + \frac{\Lambda/(2\Gamma)}{\left(\Lambda/(2\Gamma)\right)^2 + 1}\right].
\end{equation}

\subsubsection*{Conductivity in the infinite-bandwidth limit}

\label{sec:Transport-regimes}

For the separation of scales $\Gamma(\omega)\ll\Lambda$, excitations are sharply peaked such that $A^{2}(0,\xi)\approx 2\pi \delta(\xi+\Sigma_R^{\prime}(0))/\Gamma$. For an infinite bandwidth at low temperatures, where we assume $|\Sigma_R^{\prime}(0)|\ll \Lambda/2$, we obtain
\begin{equation}
\sigma = \frac{e^2}{\Gamma}\int_{-\Lambda/2}^{\Lambda/2}(\xi+\Lambda/2)\delta(\xi+\Sigma_R^{\prime}(0))\frac{d \xi}{2\pi} \approx \frac{e^2}{4\pi}\frac{\Lambda}{\Gamma}.
\label{eq:conductivity-inf-bw}
\end{equation}
For the model presently considered, $\Sigma_R^{\prime}(0)$ is identically zero in the infinite bandwidth limit. By equating the electronic bandwidth with the Fermi-energy $\Lambda/2\approx\varepsilon_{F}=k_F^2/(2m) = \pi\nu v_F^2$, and inserting $\Gamma=1/(2\tau)$, the conductivity reduces to the Drude form: $\sigma = e^2\nu v_F^2 \tau\equiv ne^2\tau/m$, where $n=m\nu v_F^2$ is the fermion particle density in two dimensions. 

If the electronic scattering rate is frequency dependent and is not a significantly greater energy scale than temperature (e.g., inelastic scattering), then in the infinite-bandwidth limit the conductivity is instead given by
\begin{equation}
\sigma =\frac{e^2\Lambda}{4\pi}\bigintssss_{-\infty}^{\infty} \left(-\frac{\partial n_F}{\partial\omega}\right)\frac{1}{\Gamma(\omega)}d\omega. 
\label{eq:conductivity-inf-2}
\end{equation}

\subsubsection*{Finite bandwidth}

For the separation of scales $T, \Lambda \ll \Gamma(\omega)$, long-lived particle-like excitations do not exist. Rather, this regime characterizes a strongly incoherent metal (bad metal), where $A(0,\xi)\approx 2/\Gamma$ for $\xi\in[-\Lambda/2,\Lambda/2]$. Assuming $|\Sigma_R^{\prime\prime}(0)|\gg |\Sigma_R^{\prime}(0)|$, the conductivity is then given by
\begin{equation}
\sigma_{\BM} = \frac{e^2}{2\pi^2}\left(\frac{\Lambda}{\Gamma}\right)^2.
\label{eq:non-boltzmann-transport}
\end{equation}

In terms of the lifetime $\tau$, the conductivity of the strongly incoherent metal is $\sigma_{\BM} = 2(e\Lambda\tau/\pi)^2$, for $\Lambda\tau\ll1$. In this regime, the resistivity necessarily exceeds the quantum unit $h/e^2$, thus describing a bad metal. This transport formula can be understood by its comparison with the standard Drude result, Eq.~\eqref{eq:conductivity-inf-bw}, which assumes that the electronic bandwidth far exceeds the inverse lifetime: $\Lambda\tau\gg1$. In this regime, the product of spectral functions is $A^2(0,\xi)\sim \tau\delta(\xi)$, which leads to the factor of $\tau$ in the conductivity, and restricts the region of phase space contributing to transport such that the velocity operator takes on its value at the Fermi-surface $v_x^2\simeq v_F^2/2$ (the factor of two arises from angular integration). However, for strongly incoherent transport ($\Lambda\tau\ll1$), the spectral function is $A(0,\xi)\sim \tau^2$, and the entire electronic band contributes to transport. 

For a large, frequency-dependent scattering rate $\Gamma(\omega)\gg\Lambda$ at intermediate temperatures, the conductivity is 
\begin{equation}
\sigma_{\BM} =\frac{1}{2}\left(\frac{e\Lambda}{\pi}\right)^2\bigintssss_{-\infty}^{\infty} \left(-\frac{\partial n_F}{\partial\omega}\right)
\left(\frac{\Gamma(\omega)}{\omega^2+\Gamma^2(\omega)}\right)^{2}d\omega\approx\frac{1}{2}\left(\frac{e\Lambda}{\pi}\right)^2\bigintssss_{-\infty}^{\infty} \left(-\frac{\partial n_F}{\partial\omega}\right)\frac{1}{\Gamma^2(\omega)} d\omega.
\label{eq:conductivity-inf-2}
\end{equation}
The lifetime $\tau\equiv1/(2\Gamma)$ in Eq.~\eqref{eq:non-boltzmann-transport} generalizes to: $\tau^2\rightarrow\overline{\tau^2}\equiv \int (-n_F^\prime(\omega)) \tau^2(\omega) d\omega$.

\subsubsection*{Ultra-high temperatures - Finite bandwidth}

For the separation of scales $\Lambda \ll\Gamma(\omega)\ll T$, the spectral function remains broad and the thermal distribution function is uniform and can be expanded in powers of inverse temperature: $-n^{\prime}_F(\omega)\sim1/(4T)$. The electrical conductivity is no longer determined solely in terms of the IR values of $\Sigma_R(\omega)$, instead the frequency dependence of the spectral function is important. Using Eq.~\eqref{eq:conductivity-arb-bw}, an estimate of conductivity in this regime is given as follows
\begin{align}
\sigma_{\TIM} &= \frac{e^2}{T}\frac{1}{(4\pi)^2}\int_{-\infty}^{\infty} \left\{\frac{(\Lambda+2\omega)}{\Gamma(\omega)}\left[\arctan\left(\frac{\Lambda-2\omega}{2\Gamma(\omega)}\right)+\arctan\left(\frac{\Lambda+2\omega}{2\Gamma(\omega)}\right)\right]+\frac{4\Lambda(\Lambda-2\omega)}{(\Lambda-2\omega)^2+4\Gamma(\omega)^2}\right\} d\omega \nonumber\\
& \approx \frac{e^2}{16\pi}\frac{\Lambda^2}{T\Gamma}.
\label{eq:conductivity-tim}
\end{align}
The final expression is valid to leading order in $\Lambda$ for a Lorentzian spectral function with a constant $\Gamma(\omega)$. We refer to the regime $\Lambda \ll\Gamma(\omega)\ll T$ as a thermal incoherent metal, as the high-temperature behavior of the distribution function determines the conductivity.

\subsection{Strange-metal regime}
\label{subsec:MFL}

In the vicinity of the QCP, the (2+1)d Y-SYK model is a Marginal Fermi liquid (MFL)~\cite{Varma1989} at low temperatures and weak coupling in the infinite bandwidth limit~\cite{Patel2023} which we use as a model of the strange-metal (SM) regime~\cite{Patel2023}. We verified this result in Eqs.~\eqref{eq:MFL-SE1} and \eqref{eq:MFL-SE2} (in the overdamped and underdamped regimes, respectively) by computing the zero-temperature fermionic self-energy on the imaginary frequency axis. On the real frequency axis at zero temperature, the fermionic self-energy in the SM regime has the form
\begin{equation}
\lim_{T\rightarrow0}\Sigma_R(\omega) \simeq \lambda_{1,2}\omega\left[\log\left(\frac{|\omega|}{\omega_c} \right)-i\frac{\pi}{2}\text{sgn}(\omega)\right] -i\Gamma_0.
\end{equation}
The MFL UV cutoff is denoted by $\omega_c$ and the potential disorder scattering rate is $\Gamma_0\equiv v^2\pi\nu$. For $|\omega|\ll \lambda_E\omega_D$, $\omega_c = e\varepsilon_{F}/(2\pi\lambda)$ and $\lambda_1 =\lambda$. For $2\pi \lambda_E\omega_D\ll|\omega|\ll\omega_{D}$, $\omega_c = e\omega_D$ and $\lambda_2 = 2\lambda$. At low temperatures and weak coupling, the static part of the Yukawa-SYK interaction contribution to the fermionic self energy in Eq.~\eqref{eq:Exact_elec_SE} can be deduced. Let $x\equiv M^2/(4\pi^2 T \omega_D \lambda_E)$. In terms of the temperature and coupling-dependent function $\Gamma_1$: 
\begin{align}
\Sigma_R^{\prime\prime}(0,T) &\simeq -(\Gamma_1(T) + \Gamma_0),  \\
\Gamma_{1}(T) &= \pi\lambda T\begin{cases}
   -\log\left[x(1+x)_{-1/2}^2\right] ;\quad T/\omega_D,M/\omega_D\ll \lambda_E\ll1\\
   2\log\left[\coth(M/(2T))\right];\quad \lambda_E\ll T/\omega_D,M/\omega_D\ll1.
\end{cases} 
\label{eq:Gamma1}
\end{align}
The argument of the logarithm has asymptotic values $x(1+x)^2_{-1/2} = \pi x +\mathcal{O}(x^2)$ and $x(1+x)^2_{-1/2} = 1-1/(4x)+\mathcal{O}(x^{-2})$.

Using Eq.~\eqref{eq:conductivity-inf-bw}, we find the low-temperature ($T\ll\Lambda,|\Sigma_R^{\prime\prime}|$) DC conductivity in the SM regime is
\begin{equation}
\label{eq:conductivity-MFL}
\sigma_{\SM}(T) \approx \frac{1}{2}\frac{e^2v_F^2\nu}{(\Gamma_{1}+\Gamma_0)}.
\end{equation}
As shown in Eq.~\eqref{eq:Gamma1}, $\Gamma_{1}(T)$ is linear in $T$, up to logarithmic corrections from the renormalized boson mass,  which is consistent with a $T$-linear resistivity~\cite{Patel2023}. For weak coupling and weak disorder, the separation of scales $T\ll|\Sigma_R^{\prime\prime}|$ cannot be established. In this case, inelastic scattering described by $\Sigma_R^{\prime\prime}(|\omega|>0)$ must be considered for a precise low-temperature transport calculation.

\subsection{Bad-metal regime}

In this section, we use the spectral function obtained in Eq.~\eqref{eq:FBW-spectral-function} to show that the (2+1) dimensional Y-SYK model exhibits strongly incoherent metallic behavior at strong coupling in a finite-bandwidth scenario, where the separation of scales $|i\omega_n-\Sigma(i\omega_n)| \gg \Lambda/2$ is satisfied. In particular, we show that the impurity regime of the (2+1)-dimensional model leads to transport consistent with bad-metal physics. Using the spectral function defined in Eq.~\eqref{eq:FBW-spectral-function}, the longitudinal conductivity in the impurity regime is given by
\begin{align}
\sigma(T) &= e^2\int_{-\Omega_0}^{\Omega_0}\delta(\omega)\left\{\int_{-\Lambda/2}^{\Lambda/2}(\xi+\Lambda/2)A(\omega,\xi)^2 \frac{d\xi}{2\pi}\right\}\frac{d\omega}{2\pi} \nonumber\\
 &= \frac{e^2}{2\pi}\int_{-\Lambda/2}^{\Lambda/2}(\xi+\Lambda/2)A^2(0,\xi) \frac{d\xi}{2\pi} \nonumber\\ 
 &= \frac{e^2}{2\pi}\int_{-\Lambda/2}^{\Lambda/2}(\xi+\Lambda/2)\left[\frac{4\Omega_0}{(2\xi)^2 + \Omega_0^2}\right]^2 \frac{d\xi}{2\pi} \nonumber\\
 &\approx 2\left(\frac{e \Lambda}{\pi \Omega_0}\right)^2; \quad T, \Lambda\ll\Omega_{0}.
 \end{align}
Note that the above result agrees with the non-Boltzmann transport formula obtained in Eq.~\eqref{eq:non-boltzmann-transport}; the reason for this agreement is due to the separation of scales $\Lambda\ll\Omega_{0}$ being valid and thus the spectral function in Eq.~\eqref{eq:FBW-spectral-function} can be approximated as just $A(\omega,\xi)\approx4/\Omega_{0}$. Therefore, using Eq.~\eqref{eq:non-boltzmann-transport}, we find
\begin{equation}
\sigma_{\BM}(T) \approx 2\left(\frac{e \Lambda}{\pi \Omega_0}\right)^2 = \frac{3^2}{2^9}\left(\frac{e\kappa}{\lambda}\right)^2; \quad T, \Lambda\ll\Omega_{0}. 
\end{equation}
Under the assumption of the separation of scales $|\Sigma_R^{\prime\prime}| \gg \Lambda$, we arrive at a universal upper bound $\sigma_{\BM}\ll 2e^2/\pi^2$, which necessarily leads to a resistivity that exceeds the Mott-Ioffe-Regel (MIR) bound~\cite{Mott1972, Hussey2004}. 

For the ultra-high-temperature limit $T\gg\Omega_0$, Eq.~\eqref{eq:conductivity-tim} gives the conductivity in the case where the lifetime is frequency independent. However, to calculate the conductivity in the impurity regime, we must use the non-Lorentzian spectral function in Eq.~\eqref{eq:FBW-spectral-function}. For the separation of scales $\Lambda\ll\Omega_{0}\ll T$, and to leading order in powers of $\Lambda/\Omega_0$, the conductivity is 
\begin{align}
\sigma_{\TIM}(T) & = 
\frac{e^2}{4T}\int_{-\Omega_0}^{\Omega_0}\left[\int_{-\Lambda/2}^{\Lambda/2}(\xi+\Lambda/2)A^2(\omega,\xi) \frac{d\xi}{2\pi}\right]\frac{d\omega}{2\pi} \nonumber\\
& = \frac{e^2}{4T}\int_{-\Omega_0}^{\Omega_0}\left[\int_{-\Lambda/2}^{\Lambda/2}(\xi+\Lambda/2)\left(\frac{4\sqrt{\Omega_0^2-\omega^2}}{(2\xi)^2-4\omega \xi + \Omega_0^2}\right)^2 \frac{d\xi}{2\pi}\right]\frac{d\omega}{2\pi} \nonumber\\
& \approx \frac{2e^2\Lambda^2}{3\pi^2 T \Omega_0}=\frac{e^2}{3\pi^2}\left(\frac{\Lambda}{T}\right)^{3/2}\frac{1}{\sqrt{\lambda\log(1+\omega_D^2/M_0^2)}}; \quad \Lambda\ll\Omega_{0}\ll T. 
\end{align}
Thus, for a wide range of temperatures, the impurity-regime solutions obtained in the strong-coupling limit using a finite electronic bandwidth yield a resistivity that exceeds the quantum unit. Although both the low-to-intermediate and ultra-high temperature regimes have resistivity greater than the quantum unit, we refer to the regime where $T\ll\Omega_0$ as the bad metal (BM) and the regime where $T\gg\Omega_0$ as the thermal incoherent metal (TIM).

\section{Shear viscosity to entropy density ratio}

\subsection{Kubo formula and spectral representation}

The real part of the shear viscosity, that is, the dissipative part, is defined by the following Kubo formula~\cite{Kadanoff1963, Hosoya1984, Ge2020}:
\begin{align} \nonumber
\eta &= -\lim_{\Omega \rightarrow 0}\frac{1}{\Omega}\text{Im}G^{xy,xy}_R(\Omega,0),\\
G^{xy,xy}_R(\Omega,0) &= i \int_{0}^{+\infty} d t \int e^{i \Omega t} \left\langle\left[\hat{T}_{x y}(\bm{r}, t), \hat{T}_{x y}(\bm{0}, 0)\right]\right\rangle d^2r.
\end{align}
The components of the energy-momentum tensor are given by $\hat{T}_{\alpha\beta}$. The complete Kubo formula for the complex shear viscosity has an additional term related to the adiabatic compressibility and an additional contact term (analogous to the diamagnetic part of the electrical conductivity~\cite{Luttinger1964, Taylor2010, Bradlyn2012, He2014}. Such terms do not contribute to the real part of the shear viscosity computed in the appropriate local limit. The Ward-Takahashi identity for the energy momentum vertex is derived in Ref.~\cite{He2014}. In particular, since the energy-momentum tensor is not uniquely defined, we can use the equations of motion to impose that it is a symmetric tensor. Doing so leads to the canonical form of this tensor. Using the Matsubara formalism, we express the Fourier transform of the time-ordered correlation function as
\begin{equation}
G^{xy,xy}(i\Omega_m,0) = 2T\sum_{\omega_n}\int \left(\frac{k_x k_y}{m}\right)^2 G(i\omega_n, \bm{k}) G(i\omega_n+i\Omega_m, \bm{k})\frac{d^2k}{(2\pi)^2}.
\end{equation}
Using the spectral representation $G(z) = \int_{\mathbb{R}} \frac{A(\omega)}{\omega-z}\frac{d\omega}{2\pi}$, where $\text{Im}z\neq0$ and $A(\omega) = -2 \text{Im}G_R(\omega)$, we obtain
\begin{equation}
G^{xy,xy}(i\Omega_m,0) = 2T\sum_{\omega_n}\int \biggl[\int \left(\frac{k_x k_y}{m}\right)^2 \frac{A(\omega, \bm{k})}{i \omega_n -\omega}\frac{A(\omega^\prime , \bm{k})}{i \omega_n + i \Omega_m - \omega^\prime}\frac{d^2 k}{(2\pi)^2}\biggr] \frac{d \omega d \omega^\prime}{(2\pi)^2}.
\end{equation}
Evaluating the Matsubara sum then performing the analytical continuation of the external frequency $i\Omega_m \rightarrow \Omega+i0^+$ leads to
\begin{equation}
G^{xy,xy}_R(\Omega,0) = 2\int \left[\int \left(\frac{k_x k_y}{m}\right)^2 A(\omega, \bm{k})A(\omega^\prime, \bm{k})\frac{n_F(\omega)-n_F(\omega^\prime )}{\omega-\omega^\prime+\Omega+i0^+}\frac{d^2 k}{(2\pi)^2}\right]\frac{d \omega d \omega^\prime}{(2\pi)^2}.
\end{equation}
Using the Sokhotski–Plemelj theorem and the real-valuedness of the spectral function, we obtain
\begin{align}\nonumber
\text{Im}G^{xy,xy}_R(\Omega,0) &=-\int \frac{d \omega d \omega^\prime}{2\pi}\int \left(\frac{k_x k_y}{m}\right)^2 A(\omega, \bm{k})A(\omega^\prime, \bm{k})\left(n_F(\omega)-n_F(\omega^\prime)\right)\delta(\omega-\omega^\prime+\Omega) \frac{d^2 k}{(2\pi)^2}\\ 
&=-\int \frac{d \omega}{2\pi}\int \left(\frac{k_x k_y}{m}\right)^2 A(\omega, \bm{k})A(\omega+\Omega, \bm{k})\left(n_F(\omega)-n_F(\omega + \Omega)\right) \frac{d^2 k}{(2\pi)^2}.
\end{align}

In terms of the spectral function, the shear viscosity is then given by
\begin{equation}
\eta =\int \left[\int \left(\frac{k_x k_y}{m}\right)^2 \left(-\frac{\partial n_F}{\partial\omega}\right) A(\omega, \bm{k})^2 \frac{d^2k}{(2\pi)^2}\right] \frac{d \omega}{2\pi}.
\label{eq:viscosity-intermediate}
\end{equation}
As in Eq.~\eqref{eq:AngularInt}, for a function $F(k)$, the angular integration is 
\begin{equation}
\int\left(\frac{k_x k_y}{m}\right)^2 F(k) \frac{d^2k}{(2\pi)^2} = \frac{1}{(2\pi m)^2}\int_{-\Lambda/2}^{\Lambda/2} k^5 F(k)\ dk\int_{0}^{2\pi}\cos^2\theta \sin^2\theta\ d\theta = \frac{\nu}{2}\int_{-\Lambda/2}^{\Lambda/2}\left(\xi+\Lambda/2\right)^2F(\xi) d\xi,
\end{equation}
where $\nu=m/(2\pi)$ is the 2d density of states per spin. The shear viscosity for the isotropic dispersion with symmetric bandwidth is
\begin{equation} 
\eta = \frac{\nu}{2}\bigintssss_{-\infty}^{\infty} \left[\bigintssss_{-\Lambda/2}^{\Lambda/2}\left(\xi+\Lambda/2\right)^2 A^2(\omega, \xi) d\xi\right] \left(-\frac{\partial n_F}{\partial\omega}\right) \frac{d \omega}{2\pi}.
\label{eq:viscosity-arb-bw}
\end{equation}

\subsection{Limiting formulae}

Evaluating Eq.~\eqref{eq:viscosity-arb-bw} in the infinite bandwidth limit, we obtain 
\begin{align}
\eta(T) &= \frac{\nu}{2}\int_{-\infty}^{\infty}\left(-\frac{\partial n_F}{\partial \omega}\right)\frac{1}{{|\Sigma_R^{\prime\prime}(\omega)|}}\left[\int_{-\Lambda/2}^{\Lambda/2} \left(\xi+\Lambda/2\right)^2\delta(\xi - \omega + \Sigma_R^{\prime}(\omega)) d\xi\right] d\omega \nonumber \\
&= \frac{\nu}{2}\int_{-\infty}^{\infty} \left(-\frac{\partial n_F}{\partial \omega}\right)\frac{1}{{{|\Sigma_R^{\prime\prime}(\omega)|}}}\left(\omega - \Sigma_R^{\prime}(\omega)+\Lambda/2\right)^2 d\omega \nonumber \\
&= \frac{\nu\Lambda^2}{8}\int_{-\infty}^{\infty} \left(-\frac{\partial n_F}{\partial \omega}\right)\frac{1}{|\Sigma_R^{\prime\prime}(\omega)|} d\omega.
\label{eq:viscosity-inf-bw-1}
\end{align}
Equation~\eqref{eq:viscosity-inf-bw-1} is evaluated numerically for the strange-metal regime in Figures 3 and 4 of the main text. The infinite-bandwidth limit is defined by $|\Sigma_R^{\prime\prime}|,|\Sigma_R^{\prime}|,T\ll \Lambda$, which results in the following lower bound for a fermion in 2d with an isotropic quadratic dispersion with symmetric bandwidth:
\begin{equation}
\eta  \gg \nu\Lambda.
\end{equation}
Dividing by the fermionic contribution to the entropy density, $s_F=2\pi^2\nu T/3$, and demanding the separation of scales $\Lambda \gg T$, we obtain a generic shear viscosity to entropy density lower bound 
\begin{equation}
\frac{\eta}{s_{F}} \gg 1.
\end{equation}
The Kovtun-Son-Starinets bound is trivially satisfied in the infinite-bandwidth limit, as expected in this regime~\cite{Kovtun2003}. Setting $\Lambda/2=\varepsilon_{F}$ in Eq.~\eqref{eq:viscosity-inf-bw-1} and approximating $\Sigma_{R}^{\prime}$ by its zero-frequency value, the shear viscosity becomes $\eta\approx\nu\varepsilon_{F}^2/(2|\Sigma^{\prime\prime}_R(0)|)\approx\nu\varepsilon_{F}^2\tau$ as expected for Boltzmann transport.

\subsection{Strange-metal regime}

In the marginal Fermi liquid regime, $|\Sigma^{\prime\prime}_R(0,T)|\approx (\Gamma_{1} + \pi v^2\nu)$, and thus $\eta(T)\approx \nu\varepsilon_{F}^2/(2(\Gamma_1 + \pi v^2\nu))$, where we have set $\Lambda/2=\varepsilon_F$. At low temperatures, the free-boson contribution to the entropy $s_{B,0} \propto T^2$ is negligible compared to the the free-fermion and interaction contributions: $s_{F}\approx2\pi^2\nu T/3$ and $s_{\text{Int}}\approx2\pi^2\lambda\nu T L(0)/3$, respectively. The low-temperature shear viscosity to entropy density ratio for intermediate values of coupling is thus given by
\begin{equation}
(\eta/s)_{\SM} \approx \frac{3}{4\pi^2}\left(\frac{\varepsilon_{F}}{T}\right)\frac{\varepsilon_{F}}{(\Gamma_{1}+\Gamma_0)(1+\lambda L(0))}.
\end{equation}
For $\lambda_E\ll T/\omega_D,M/\omega_D\ll1$, $\Gamma_1(T)$ is given by the second line of Eq.~\eqref{eq:Gamma1} with $M/T \simeq \text{arcosh}(3/2)$, such that $\Gamma_1\simeq\pi\log(5)\lambda T$. In the absence of spatially random potential disorder, and for weak values of coupling such that $\lambda L(0) \ll 1$, we obtain
\begin{equation}
\label{eq:eta-s-MFL-weakestcoupling}
(\eta/s)_{\SM} \approx \frac{3}{4\pi^3\log(5)}\left(\frac{\varepsilon_{F}}{T}\right)^2\frac{1}{\lambda}.
\end{equation}
Equation~\eqref{eq:eta-s-MFL-weakestcoupling} is the weak-coupling limit appearing in Fig.~(5) of the main text. In the presence of nonzero potential disorder, in the weak-coupling limit, $\Gamma_{1}\ll\Gamma_{0}$, and thus $(\eta/s)\approx(3/4\pi^2)(\varepsilon_{F}/\Gamma_0)(\varepsilon_F/T)$ for $\Gamma_0,T\ll\varepsilon_F$.

\subsection{Bad-metal regime}

For the separation of scales $|\Sigma_R^{\prime\prime}(\omega)| \gg \Lambda,T,|\Sigma_R^{\prime}(\omega)|$, $A(\omega,\xi)\approx 2/|\Sigma_R^{\prime\prime}(\omega)|$, then
\begin{align}
\eta(T) &= \frac{\nu}{\pi} \int_{-\infty}^{\infty} \left(-\frac{\partial n_F}{d\omega}\right)\frac{1}{\Sigma_R^{\prime\prime}(\omega)^2} \left[\int_{-\frac{\Lambda}{2}}^{\frac{\Lambda}{2}} \left(\xi+\Lambda/2\right)^2 d\xi \right] d\omega\nonumber\\
&= \frac{\nu\Lambda^3}{3\pi}\int_{-\infty}^{\infty} \left(-\frac{\partial n_F}{d\omega}\right)\frac{1}{\Sigma^{\prime\prime}_R(\omega)^2} d\omega.
\label{eq:viscosity-finite-bw}
\end{align}
In the impurity regime, $\Sigma_R^{\prime\prime}(\omega)\approx-\Omega_0/2$, for $|\omega|\ll\Omega_0$. For temperatures $T\ll\Omega_0$, we obtain $\eta_{\BM}=\nu\Lambda^3 /(3 \pi \Sigma_R^{\prime\prime}(0)^2)=4\nu\Lambda^3/(3\pi\Omega_0^2)$. The separation of scales $|\Sigma^{\prime\prime}_{R}(\omega)|\gg\Lambda/2$ places an \emph{upper bound} on the shear viscosity: $\eta_{\BM}\ll 4\nu\Lambda/(3\pi)$. For temperatures $T\ll\Omega_0,\omega_D$, the shear viscosity to entropy density ratio becomes
\begin{equation}
(\eta/s)_{\BM}=\frac{4\Lambda^3}{3\pi\Omega_0^2}\frac{1}{2\pi \Lambda T/\Omega_0 + 6\zeta(3)\varepsilon_{F}(T/\omega_D)^2}.
\end{equation}
For sufficiently strong coupling at nonzero temperatures, $\omega_D^2\Lambda/(\varepsilon_{F}\Omega_0)\ll T$, and the free-boson contribution to the entropy density is dominant. We thus obtain an upper bound on $\eta/s$ given by
\begin{equation}
(\eta/s)_{\BM}\ll \frac{2}{9\pi\zeta(3)}\frac{\Lambda}{\varepsilon_{F}}\left(\frac{\omega_D}{T}\right)^2. 
\end{equation}
Using the result for $\Omega_{0}$ in Eq.~\eqref{eq:Omega0}, the shear viscosity to entropy density ratio, for $T\ll\omega_D$, becomes
\begin{equation}
(\eta/s)_{\BM}\simeq\frac{\pi}{128\zeta(3)}\frac{\varepsilon_{F}M_0^4}{\Lambda\omega_D^4} \left(\frac{\omega_D}{T}\right)^2\frac{1}{\lambda^{2}} = \frac{\pi\kappa}{256\zeta(3)}\left(\frac{M_0}{T}\right)^2\frac{1}{\lambda^2}. 
\end{equation}

In the ultra-high-temperature limit $T\gg\Omega_0$, the frequency dependence of $\Sigma_R(\omega)$ cannot be ignored and we must use the spectral function in Eq.~\eqref{eq:FBW-spectral-function}. For the separation of scales $\Lambda\ll\Omega_{0}\ll T$, and to leading order in powers of $\Lambda/\Omega_0$, the shear viscosity is 
\begin{align}
\eta_{\TIM}(T) & = 
\frac{\nu}{8T}\int_{-\Omega_0}^{\Omega_0}\left\{\int_{-\Lambda/2}^{\Lambda/2}(\xi+\Lambda/2)^2 A(\omega,\xi)^2 d\xi\right\}\frac{d\omega}{2\pi} \nonumber\\
& = \frac{\nu}{8T}\int_{-\Omega_0}^{\Omega_0}\left\{\int_{-\Lambda/2}^{\Lambda/2}(\xi+\Lambda/2)^2\left[\frac{4\sqrt{\Omega_0^2-\omega^2}}{(2\xi)^2-4\omega \xi + \Omega_0^2}\right]^2 d\xi\right\}\frac{d\omega}{2\pi} \nonumber\\
& \approx \frac{4\nu\Lambda^3}{9\pi T \Omega_0}=\frac{2\nu\Lambda}{9\pi}\left(\frac{\Lambda}{T}\right)^{3/2}\frac{1}{\sqrt{\lambda\log(1+\omega_D^2/M_0^2)}}; \quad \Lambda\ll\Omega_{0}\ll T. 
\end{align}

\section{Diffusivity and Transport Summary}

\subsection{Charge diffusivity in the bad-metal and thermal incoherent-metal regimes}

Here, we compute the diffusivity in the impurity regime. As delineated in Sec.~\ref{subsec:Transport lifetime}, temperatures $T\ll\Omega_0$ ($T\gg\Omega_0$) of the impurity regime correspond to a bad metal (thermal incoherent metal). The number density is
\begin{equation}
n = 2\int_{\varepsilon_{\bm{k}}\in[0,\Lambda]} \left[\int_{-\infty}^{\infty}n_F(\omega) A(\omega,\bm{k}) \frac{d\omega}{2\pi}\right]\frac{d^2\bm{k}}{(2\pi)^2}\approx 2\nu\int_{-\mu}^{\Lambda-\mu}  \left[\int_{-\Omega_0}^{0} \frac{4\sqrt{\Omega_0^2-\omega^2}}{(2\xi)^2-4\omega \xi + \Omega_0^2} \frac{d\omega}{2\pi}\right]d\xi.
\label{eq:numberdensity}
\end{equation}
The spectral function is given by Eq.~\eqref{eq:FBW-spectral-function}. On the right-hand side we have assumed $T\ll\Lambda, \Omega_0$, such that $n_F(\omega)\approx\theta(-\omega)$. Charge compressibility, $\chi\equiv\partial n/\partial\mu$, is given by
\begin{equation}
\chi_{\BM} = 2\nu \frac{\partial}{\partial \mu}\int_{-\mu}^{\Lambda-\mu}\left[\int_{-\Omega_0}^{0}A(\omega,\xi)\frac{d\omega}{2\pi}\right]d\xi = 2\nu \int_{-\Omega_0}^{0}\left[A(\omega,-\mu)-A(\omega,\Lambda-\mu)\right]\frac{d\omega}{2\pi}.
\end{equation}
In the case of a symmetric bandwidth, $\mu=\Lambda/2$, $n=\nu\Lambda$, and $\chi =\nu \int_{-\Omega_0}^{0}\left[A(\omega,-\Lambda/2)-A(\omega,\Lambda/2)\right]d\omega/\pi$. For the impurity regime, compressibility is thus given by
\begin{align}
    \chi_{\BM} &= -16\nu\Lambda \int_{-\Omega_0}^{0} \frac{\omega\sqrt{\Omega_0^2-\omega^2}}{(\Lambda^4+\Omega_0^4+2\Lambda^2(\Omega_0^2-2\omega^2))}\frac{d\omega}{\pi} \nonumber\\ 
    &= \frac{2\nu}{\pi}\left[2\Omega_0/\Lambda-\left(\left(\Omega_0/\Lambda\right)^2-1\right)\text{arctan}\left(\frac{2\Lambda\Omega_0}{\Omega_0^2-\Lambda^2}\right)\right] \nonumber\\
    &=16\nu\Lambda/(3\pi\Omega_0)+O(\Lambda^3), \quad T\ll\Omega_0.
\end{align}
For a particle-hole symmetric system, the diffusion constant is given by the Nerst-Einstein relation, $D=\sigma/(e^2\chi)$. 
Combining with conductivity $\sigma_\BM/e^2 = 2(\Lambda/(\pi\Omega_0))^2$, we obtain $D_{\BM}=3\Lambda/(8\pi\nu\Omega_0)$. In more conventional units ($\Lambda \simeq 2\varepsilon_{F} = m v_F^2$), we obtain $D_{\BM} =(3/4)v_F^2/\Omega_0$, where $\Omega_0$ is the impurity inverse lifetime in the IR of the bad metal regime. Compared with the diffusivity bound $D\gtrsim v_F^2/T$~\cite{Hartnoll2015}, and the assumed separation of scales $T\ll\Omega_0$, we find that within the impurity regime the diffusivity bound is violated. 

For $T\gg \Omega_0$, $n_F(\omega)\approx1/2-\omega/(4T)+\cdots$ in the interval $\omega\in[-\Omega_0,\Omega_0]$. The leading-order, non-vanishing contribution to the high-temperature compressibility is thus given by
\begin{align}
\chi_{\TIM} &\approx \frac{\nu}{\pi}\int_{-\Omega_0}^{\Omega_0}\left(\frac{1}{2}-\frac{\omega}{4T}\right)\left[A(\omega,-\Lambda/2)-A(\omega,\Lambda/2)\right] d\omega \nonumber\\
&\approx \frac{4\nu\Lambda}{\pi T} \int_{-\Omega_0}^{\Omega_0} \frac{\omega^2\sqrt{\Omega_0^2-\omega^2}}{(\Lambda^4+\Omega_0^4+2\Lambda^2(\Omega_0^2-2\omega^2))} d\omega = \frac{\nu\Lambda}{2T}.
\end{align}
Combining with the high-temperature conductivity, $\sigma_{\TIM} = 2e^2\Lambda^2/(3\pi^2 T\Omega_0)$, we obtain the high-temperature diffusivity $D_{\TIM}=4\Lambda/(3\pi^2\nu\Omega_0)$. Although the high-temperature diffusivity remains $\sim\Lambda/(\nu\Omega_0)$, the energy-scale has the high-temperature limit $\Omega_0\sim 2 \sqrt{\lambda\Lambda T \log(1+\omega_D^2/M_0^2)}$. Setting $\Lambda\simeq 2\varepsilon_F$, the diffusivity is given by 
\begin{equation} 
D = \frac{v_F^2}{\Omega_0}\times
\begin{cases}
3/4;\quad T\ll \Omega_0,\\ 8/(3\pi);\quad T \gg \Omega_0.
\end{cases} 
\label{eq:impurity-diffusivity}
\end{equation}
For the coupling dependence, we find $D\propto 1/\lambda$ for $T\ll\Omega_0$ and $D\propto 1/\sqrt{\lambda T}$ for $T\gg\Omega_0$ corresponding to the bad metal and thermal incoherent metal regimes, respectively. Our results also agree with the upper bound on diffusivity proposed in Ref.~\cite{Hartman2017}.

\subsection{Summary of transport formulae}\label{subsec:Transport lifetime}

In the low-temperature, infinite-bandwidth limit, the single-particle lifetime $\tau \equiv 1/(2\Gamma)\approx1/(2|\Sigma_R^{\prime\prime}(0))|$ is equivalent to the transport lifetime, defined $\tau_{\text{tr}}\equiv2\pi \sigma/(e^2\Lambda)=\tau$. In terms of transport lifetime, we obtain a Drude-like relation $\sigma = e^2\nu v_F^2\tau_{\text{tr}}$, for DC conductivity in the infinite bandwidth limit. In the finite-bandwidth limit with the separation of scales $T,\Lambda\ll \Gamma$, the relation between transport and single-particle lifetimes is
\begin{equation}
\tau_{\text{tr}} = (4\Lambda/\pi) \tau^2.
\end{equation} 
This formula illustrates the non-Boltzmann transport of a strongly incoherent metal. The low-temperature Boltzmann (Drude-like) and non-Boltzmann (bad metal) relations are universal within their respective separation of scales. At ultra-high temperatures, $\Lambda\ll\Gamma\ll T$, the transport-time relation becomes $\tau_{\text{tr}}=\Lambda \tau/(4T)$. In summary,
\begin{equation}
\sigma=e^2
\begin{cases}
\Lambda\tau/(2\pi);\quad \; \;\Lambda\tau\gg1; \;T\ll\Lambda,\tau^{-1}\;\;\;\; \text{\ Boltzmann Transport}\\
2(\Lambda\tau/\pi)^2;\; \quad \Lambda\tau\ll1; \; T, \Lambda \ll \tau^{-1}\;\quad \text{Strongly Incoherent/Bad-Metal Transport}\\
\propto\Lambda^2\tau/T;\;\quad \Lambda\tau\ll1; \; T\gg\Lambda,\tau^{-1}\quad \text{\  High-temperature/Thermal Incoherent Metal Transport}.
\end{cases}       
\end{equation}
For shear viscosity,
\begin{equation}
\eta=
\begin{cases}
\nu\Lambda^2\tau/4;\quad \quad \quad \;\Lambda\tau\gg1; \; T\ll\Lambda,\tau^{-1}\quad \text{\ Boltzmann Transport}\\
4\nu\Lambda^3\tau^2/(3\pi);\quad \Lambda\tau\ll1; \; T,\Lambda\ll\tau^{-1}\; \quad \text{Strongly Incoherent/Bad-Metal Transport}\\
\propto\nu\Lambda^3\tau/T;\quad \quad \Lambda\tau\ll1; \; T\gg\Lambda,\tau^{-1}\quad \text{\  High-temperature/Thermal Incoherent Metal Transport}.
\end{cases}       
\end{equation}
In the above expressions, the proportionality symbol is used for thermal incoherent transport because while the functional dependence can be determined from scaling analysis, the prefactor depends on the specific non-Lorentzian nature of the spectral function. 

\section{Superconducting critical temperature}

The linearized gap equation for the auxiliary function $\Phi\left(i \omega_n\right)$ in the Y-SYK model in (2+1)d reads~\cite{Gnezdilov2025}
\begin{align}\label{eq:gapeq1}
\Phi(i\omega_n)=g^2T_c\sum_{\omega_{n^\prime}}\mathcal{D}({i\omega_{n}-i\omega_{n^\prime}})\mathcal{F}(i\omega_{n^{\prime}}).
\end{align}
The linearized component of Gor’kov’s anomalous Green’s function is
\begin{equation} 
\label{eq:F-gapeq}
\mathcal{F}(i\omega_{n}) \equiv\int G\left(i \omega_n, \bm{k}\right) G\left(-i \omega_n,-\bm{k}\right) \Phi\left(i \omega_n\right) \frac{d^2 k}{(2 \pi)^2}=\nu\Phi(i\omega_n)
\begin{cases}
\pi/(|\omega_n|Z(i\omega_n));\quad\text{for } \Lambda/2 \gg |\omega_n Z(i\omega_n)|\\
\Lambda/(\omega_n Z(i\omega_n))^2;\quad\text{for } \Lambda/2\ll |\omega_n Z(i\omega_n)|,\\
\end{cases}       
\end{equation}
in terms of the even-frequency renormalization function $Z(i\omega_n)\equiv1+i\Sigma(i\omega_n)/\omega_n$~\cite{AGD}. The separation of scales $\Lambda/2 \gg |\omega_n Z(i\omega_n)|$ in the first line of Eq.~\eqref{eq:F-gapeq} defines the infinite bandwidth limit. In this limit, Ref.~\cite{Gnezdilov2025} showed that $T_c\simeq0.183\sqrt{\lambda}\omega_D$, for strong coupling and with a small adiabaticity ratio, $\omega_D/\varepsilon_{F}\ll1$, before saturating to a universal constant $T_c\approx0.041\varepsilon_{F}$ in the asymptotically strong coupling regime for an arbitrary adiabatic ratio. 

The second line of Eq.~\eqref{eq:F-gapeq} corresponds to the finite bandwidth limit. Inserting this result into the linearized gap equation and defining $\Delta(i\omega_n)\equiv\Phi(i\omega_n)/Z(i\omega_n)$, we obtain
\begin{equation}
\Delta(i\omega_n)Z(i\omega_n)=\nu\Lambda g^2 T_c \sum_{\omega_{n^\prime}} \frac{\mathcal{D}(i\omega_n-i\omega_{n^\prime})}{\omega_{n^\prime}^2Z(i\omega_{n^{\prime}})}\Delta(i\omega_{n^{\prime}}).
\end{equation}

Here, we obtain $T_c$ from the linearized gap equation in the finite bandwidth limit, and compare our results with the lower-dimensional pairing problem considered in Ref.~\cite{EsterlisSchmalian2019}. We find that the UV cutoff scales $\Lambda$, $\omega_D$, appearing in the 2d pairing problem, play a key role in setting the upper bound that $T_c$ saturates in the strong-coupling regime. Lastly, we recover the value for $T_c$ in the strong-coupling limit found in Ref.~\cite{EsterlisSchmalian2019}, for the zero-dimensional theory in the limit of small bandwidth. For strong coupling in the impurity regime $|\omega_n|Z(i\omega_n)\approx \Omega_0/2$. As in Ref.~\cite{EsterlisSchmalian2019}, after removal of the resonant term, we neglect the boson mass and self-energy since $|M_0^2-\Pi(i\Omega_m)|\ll \Omega_m^2$ near $T_c$. 

\begin{figure}[ht]
\centering\includegraphics[width=.5\columnwidth]{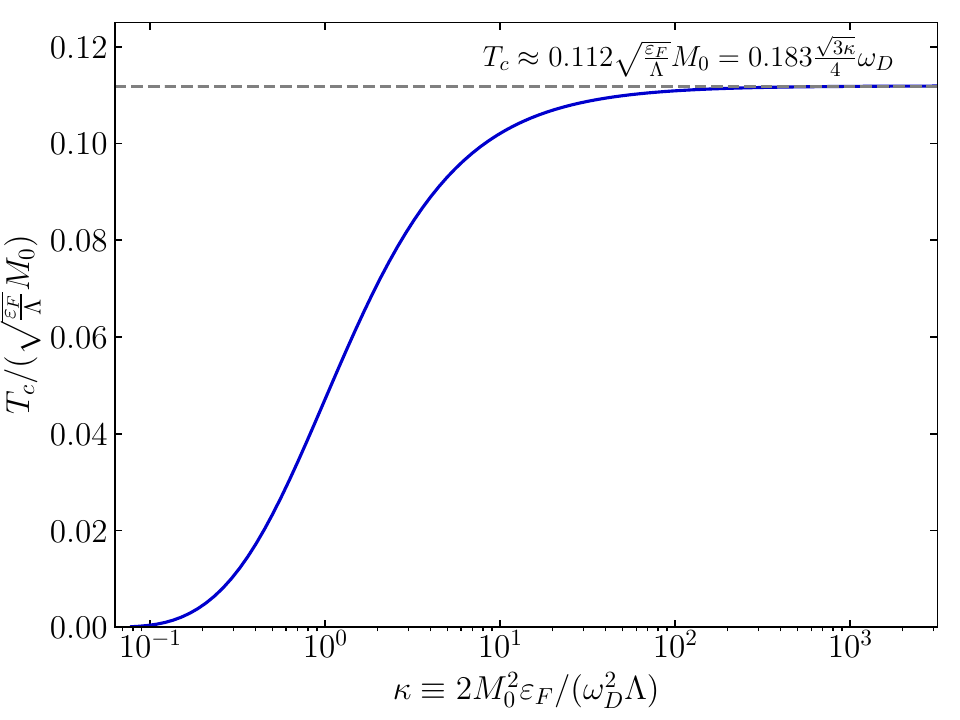}
\caption{Superconducting critical temperature in units of $\sqrt{\varepsilon_{F}/\Lambda}\;M_0$ versus the dimensionless ratio $\kappa\equiv2M_0^2\varepsilon_{F}/(\omega_D^2 \Lambda)$.}
\label{fig:Tc}
\end{figure}

For $T_c\gtrsim\omega_D$, we can expand the momentum-integrated bosonic propagator $\mathcal{D}(i\Omega_m)\approx \omega_D^2/(4\pi c^2 \Omega_m^2)$ and then recover the pairing problem in Ref.~\cite{EsterlisSchmalian2019}, modified by parameters from the (2+1)d theory. However, for $T_c\lesssim \omega_D$, the bosonic propagator cannot be expanded, and instead we must consider the full pairing problem in (2+1)d: 
\begin{equation}\label{eq:full-pairing-problem}
z(i\omega_n)\Delta(i\omega_n)=\frac{3 \kappa}{16}\sum_{n^\prime=0}^{\infty}\left[(1-\delta_{n,n^{\prime}})L(i\omega_{n}-i\omega_{n^{\prime}})+L(i\omega_{n}+i\omega_{n^{\prime}})\right]\frac{\Delta(i\omega_{n^{\prime}})}{2n^{\prime}+1},
\end{equation}
for the impurity regime. Here, the summation is given strictly in terms of positive frequencies. The logarithm is defined by $L(i\Omega_m)\equiv\log(1+\omega_D^2/\Omega_m^2)$, and we utilize the renormalization function (over positive frequencies and excluding the zero-th Bosonic Matsubara frequency, which does not contribute) defined
\begin{equation}\label{eq:zed-for-pairing-problem}
    z(i\omega_n) = 1+\frac{1}{2n+1}\frac{3\kappa }{8}\sum_{m=1}^{n}L(i\Omega_m).
\end{equation}
Equations~\eqref{eq:full-pairing-problem} and~\eqref{eq:zed-for-pairing-problem} are solved numerically and displayed in Fig.~\ref{fig:Tc}.

For large values of the dimensionless ratio $\kappa\equiv2M_0^2\varepsilon_{F}/(\omega_D^2 \Lambda)$, the superconducting critical temperature saturates to its 0d value.  In terms of the constant $c_{E}\equiv 3\kappa/16$, the superconducting critical temperature in the impurity regime is given by $T_c\simeq 0.183\sqrt{c_{E}}\omega_D$ in the limit $c_{E}\rightarrow \infty$, where $0.183\simeq0.112\times\sqrt{8/3}$~\cite{Valentinis2023b}. The limit $c_{E}\rightarrow \infty$ is equivalent to the limit $\Lambda\rightarrow 0$ with the other energy scales fixed, or the (0+1)d case, as in Ref.~\cite{EsterlisSchmalian2019}. This result reproduces the Allen-Dynes formula, where the role of the dimensionless coupling is substituted for a coupling-independent ratio of the electronic and bosonic energy scales. This correspondence may be understood by recognizing the quantum-critical pairing problem problem of the zero-dimensional model as being identical to the $\gamma=2$ case of the $\gamma-$model. In non-adiabatic units ($\Lambda/2=\varepsilon_{F}=\omega_D=M_0$), the critical temperature in the impurity regime is given by $T_c \simeq 0.047\varepsilon_{F}$, particularly close to the value in the infinite bandwidth and asymptotically strong coupling limit~\cite{Gnezdilov2025}.

The superconducting critical temperature in the $\kappa\ll1$ limit can be obtained by performing the same low-temperature weak-coupling analysis as in Ref.~\cite{Gnezdilov2025}. The pairing problem in the weak-coupling regime of the 2d Y-SYK model in the infinite bandwidth limit is identical to that of the strong-coupling (impurity) regime for a finite electronic bandwidth under the substitution $\lambda\rightarrow 3\kappa/16$. Thus, we obtain $\log(\omega_D/T_c)\sim 1/\sqrt{\kappa}$ for $\kappa\ll1$.

\bibliography{References}